\documentclass[aps,superscriptaddress,nofootinbib,floatfix,amsfonts,twocolumn]{revtex4-2}
\usepackage[bookmarksnumbered,bookmarksopen,colorlinks,citecolor=red,linkcolor=blue]{hyperref}
\usepackage{graphicx}
\usepackage{bm}
\usepackage{amsmath}
\usepackage{xcolor}
\usepackage[normalem]{ulem}
\usepackage{booktabs}
\usepackage{multirow}
\usepackage{diagbox}
\usepackage{here}

\graphicspath{{./figs/}}

\begin{document}
\title{Heavy flavour hadron production in relativistic heavy ion collisions at RHIC and LHC in EPOS4HQ}

\author{Jiaxing Zhao}
\affiliation{SUBATECH, Nantes University, IMT Atlantique, IN2P3/CNRS,
4 rue Alfred Kastler, 44307 Nantes cedex 3, France}
\author{Joerg Aichelin}
\affiliation{SUBATECH, Nantes University, IMT Atlantique, IN2P3/CNRS,
4 rue Alfred Kastler, 44307 Nantes cedex 3, France}
\author{Pol Bernard Gossiaux}
\affiliation{SUBATECH, Nantes University, IMT Atlantique, IN2P3/CNRS,
4 rue Alfred Kastler, 44307 Nantes cedex 3, France}
\author{Vitalii Ozvenchuk}
\affiliation{ Institute of Nuclear Physics, Polish Academy of Sciences, Cracow, Poland}
\author{Klaus Werner}
\affiliation{SUBATECH, Nantes University, IMT Atlantique, IN2P3/CNRS,
4 rue Alfred Kastler, 44307 Nantes cedex 3, France}

\date{\today}

\begin{abstract} 
\noindent
Employing the recently developed EPOS4HQ event generator, we study the production of different heavy-flavor mesons in relativistic heavy-ion collisions at RHIC and LHC energies. The transverse momentum spectra, yield ratio, nuclear modification factor, and elliptic flow can be well described in the EPOS4HQ framework.  We furthermore analyze the processes which modify these observables as compared to $pp$ collisions and are at the origin of the experimentally determined nuclear modification factor $R_{AA}$. 
\end{abstract}


\maketitle
\section{Introduction}
\label{sec.intro}
Theory predicts that matter at high temperature and/or high density forms a plasma of   quarks and gluons (QGP) in which partons are not confined anymore. One of the main goals of heavy-ion reactions at ultra-relativistc energies is to study this new form of deconfined matter. In the last years ample evidence has been gathered that this state exists. The observation of strangeness enhancement~\cite{STAR:2011fbd}, flow harmonics compatible with hadronization after an hydrodynamic expansions \cite{PHENIX:2006dpn}, jet quenching~\cite{CMS:2016xef}, and quarkonium suppression~\cite{Matsui:1986dk} points in this direction.

Presently the main objective is to get a more quantitative understanding and investigate the properties of the QGP matter. 

The agreement of observables like spectra and flow harmonics with viscous hydrodynamical calculations reveals the strongly-coupled nature of the QGP.  The very small viscosity $\eta/s$~\cite{Teaney:2003kp},  which is necessary to bring the calculations in agreement with data,  is close to the lowest possible  limit given by AdS/CFT~\cite{Kovtun:2004de}. So we have strong evidence that a system has been created that behaves as an almost ideal fluid.

Despite the fact that we have a good qualitative understanding about the space-time evolution of the system, there are many open questions when it comes to a quantitative analysis. The above-mentioned probes (yields, spectra, flow harmonics) give only an indirect access to the properties of expanding system, one has to rely on model calculations or simulations -- which have uncertainties. So we need additional probes to get complementary information about the system, done in the same model / simulation.  Heavy flavor hadrons have turned out to be an almost ideal probe to study the time evolution of the QGP due to the following reasons: a) The heavy quark mass ($m_c=1.5\rm GeV$ and $m_b=4.5\rm GeV$) is much larger than the QCD cutoff, $\Lambda_{QCD}\approx 200\rm MeV$. Therefore their production can be well described by perturbative QCD (pQCD). b) Heavy quarks are produced at the early stage of a heavy ion collisions and witness all later stages of the collision; c) The heavy quark mass is much larger than the typical temperature of the QGP medium, which is about a couple of hundreds of MeV, estimated by the spectrum of the directly produced  photons~\cite{ALICE:2022wpn}. Consequently,  the mass of heavy quark changes little in the hot medium and their number is conserved during the evolution. 

With these advantages, heavy flavor physics has attracted a lot of attention from both the theoretical and experimental communities and several models have been advanced to describe the heavy flavor observables.
The PHSD \cite{Song:2015sfa,Song:2015ykw} approach, in which the heavy quark physics is embedded in an approach which described the light hadrons as well, is based in the dynamical quasi particle model (DQPM), which respects the equation of state of strongly interacting matter. The Catania model \cite{Plumari:2017ntm,Scardina:2017ipo,Minissale:2020bif} is also based on
a DQPM approach and describes the expanding medium by a Boltzmann equation. Based on the EPOS3 event generator we have also studied heavy quark production \cite{Nahrgang:2014vza}
using a elastic and inelastic pQCD cross sections. The LBT model \cite{Cao:2016gvr} solves as well a Boltzmann equation for the heavy quarks including elastic and inelastic collisions but the medium is described by viscous hydrodynamics. Other models like TAMU\cite{He:2019vgs,He:2019tik}, Duke\cite{Cao:2015hia} and Torino \cite{Beraudo:2022dpz,Beraudo:2023nlq} use a Fokker Planck equation to describe the dynamics of heavy quarks and model model the expanding QGP by ideal or viscous hydrodynamics. Apart from PHSD and EPOS all these models have in common that they concentrate on heavy quark physics only and do not take advantage of the fact that the many light hadron observables allow for assessing the expansion of the QGP despite  different expansion properties of the QGP may have a strong influence on the heavy quark observables~\cite{Gossiaux:2011ea}.

Thanks to the high statistic data, which are now or become soon available, the error of the key observables, such as the enhancement of the baryon to meson ratio and the elliptic flow, are strongly reduced. This allows for a detailed quantitative comparison  between theory and experiment. In this paper we compare the heavy hadron observables with the results of EPOS4HQ. EPOS4HQ is the heavy hadron extension of the recently advanced EPOS4 approach \cite{werner:2023-epos4-overview,werner:2023-epos4-heavy,werner:2023-epos4-smatrix,werner:2023-epos4-micro}, which has been successfully used to study the light hadron observables. In the EPOS4HQ approach the heavy flavor production has been substantially improved in comparison with the former EPOSLHC approach.  Heavy flavor quarks can now be produced in hard process, as well as by gluon splitting and flavour excitation. The interaction of heavy partons with the QGP includes elastic and gluon emission reactions refs.~\cite{Gossiaux:2008jv,Aichelin:2013mra}.  After the hadronization the heavy hadrons still have final state interactions, modeled by UrQMD. The calculations are based on version EPOS4.0.1.s9.

In this paper we will systematically investigate all published heavy favor observables.
It is organized as follows. The heavy quark initial production and the medium evolution in EPOS4HQ are discussed in Section~\ref{sec.epos4}. Next, the heavy quark initial spectra and energy loss mechanism is subject of   Sections~\ref{sec.initial} and~\ref{sec.epos4hq}. In Section~\ref{sec.hadronization}, we will present the hadronization  in EPOS4HQ, which differs substantially from former heavy quark calculations with EPOS2/EPOS3/EPOSLHC by including baryons and excited mesons states. The results and the comparison with experimental data are shown in Section~\ref{sec.results}. A conclusion will be given in Section~\ref{sec.summary}.

\section{EPOS4}
\label{sec.epos4}
\subsection{EPOS4 primary interactions}

A fundamental ingredient of the EPOS4 approach \cite{werner:2023-epos4-overview,werner:2023-epos4-heavy,werner:2023-epos4-smatrix,werner:2023-epos4-micro}
is the observation that multiple partonic scatterings must strictly
happen in parallel, and not sequentially, based on very elementary
considerations concerning time-scales. To take this into account,
EPOS4 brings together ancient knowledge about S-matrix theory (to
deal with parallel scatterings) and modern concepts of perturbative
QCD and saturation, going much beyond the usual factorization approach.
The parallel scattering principle requires sophisticated Monte Carlo
techniques, inspired by those used in statistical physics to investigate
the Ising model.

In the EPOS4 approach, we distinguish ``primary scatterings\textquotedblright{}
and ``secondary scatterings\textquotedblright . The former refer
to the above-mentioned parallel scatterings with the initial nucleons
(and their partonic constituents) being involved, happening at very
high energies instantaneously. The theoretical tool is S-matrix theory,
using a particular form of the proton-proton scattering S-matrix (
``classical'' Gribov-Regge approach \cite{Gribov:1967vfb,Gribov:1968jf,GribovLipatov:1972,Abramovsky:1973fm}).
Within such an approach, one can deduce the very important AGK theorem
\cite{Abramovsky:1973fm}, which leads to factorization and binary
scaling in nuclear scatterings, which is not trivial in a multiple
scattering scheme. However, introducing energy-momentum sharing \cite{Drescher:2000ha}
(which is absolutely crucial for realistic event-by-event simulations,
AGK is violated (and so is factorization and binary scaling). The
main new development in EPOS4 \cite{werner:2023-epos4-overview,werner:2023-epos4-heavy,werner:2023-epos4-smatrix,werner:2023-epos4-micro}
is a way to accomodate simultaneously: (1) rigorous parallel scattering,
(2) energy-momentum sharing, (3) AGK theorem and factorization for
hard processes, by introducing (in a very particular way) saturation,
compatible with recent ``low-x-physics'' considerations \cite{Gribov:1983ivg,McLerran:1993ni,McLerran:1993ka,kov95,kov96,kov97a,jal97,jal97a,kov98,kra98,jal99a,jal99b}. 

\begin{figure}[h]
\centering{}\includegraphics[scale=0.22]{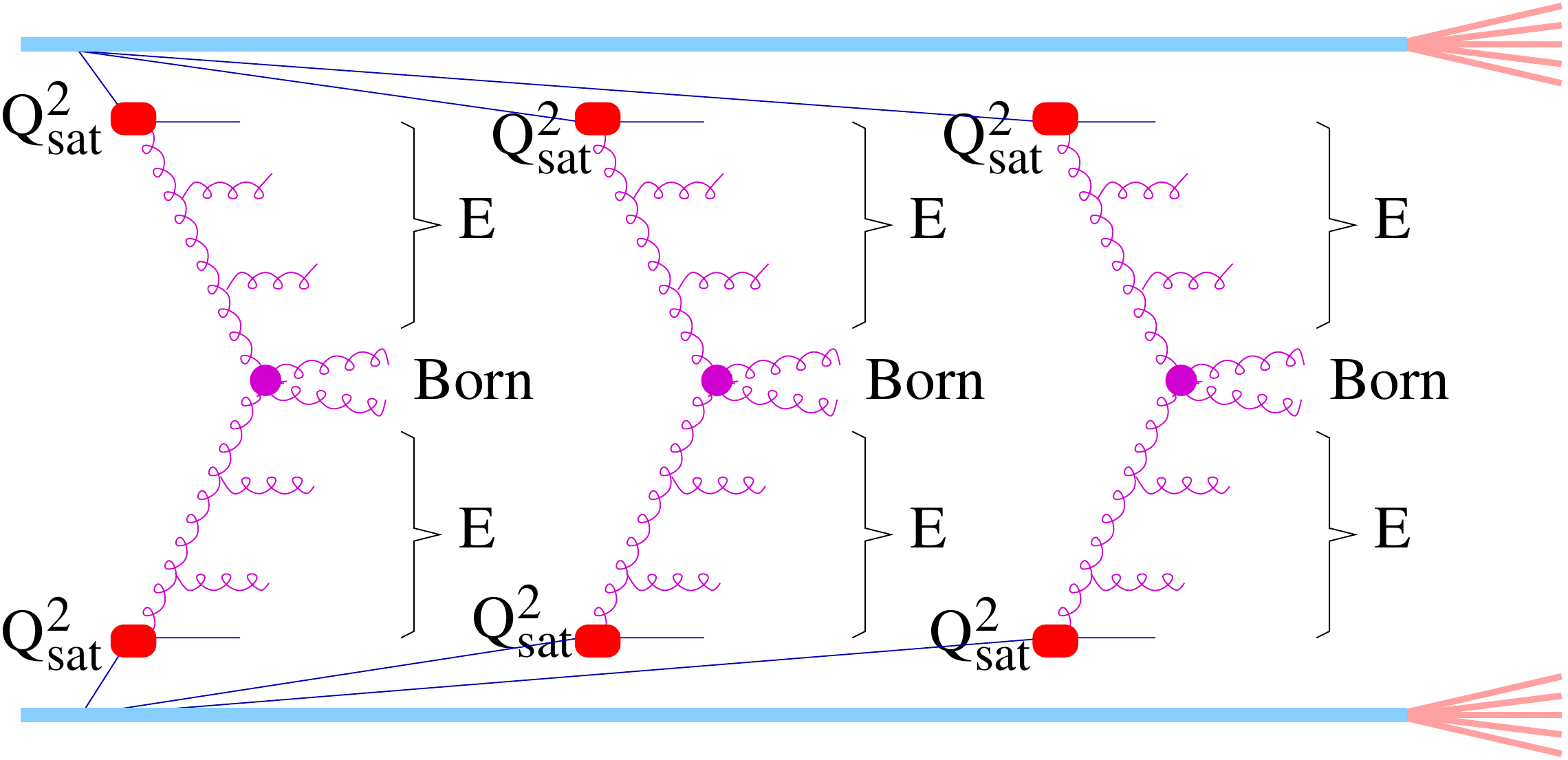} 
\caption{Rigorous parallel scattering scenario, for $n=3$ parallel scatterings,
including non-linear effects via saturation scales. The red symbols
should remind us that the parts of the diagram representing nonlinear
effects are replaced by simply using saturation scales. \label{saturation-three-pom}}
\end{figure}

\begin{figure}[h]
\centering{}\includegraphics[scale=0.22]{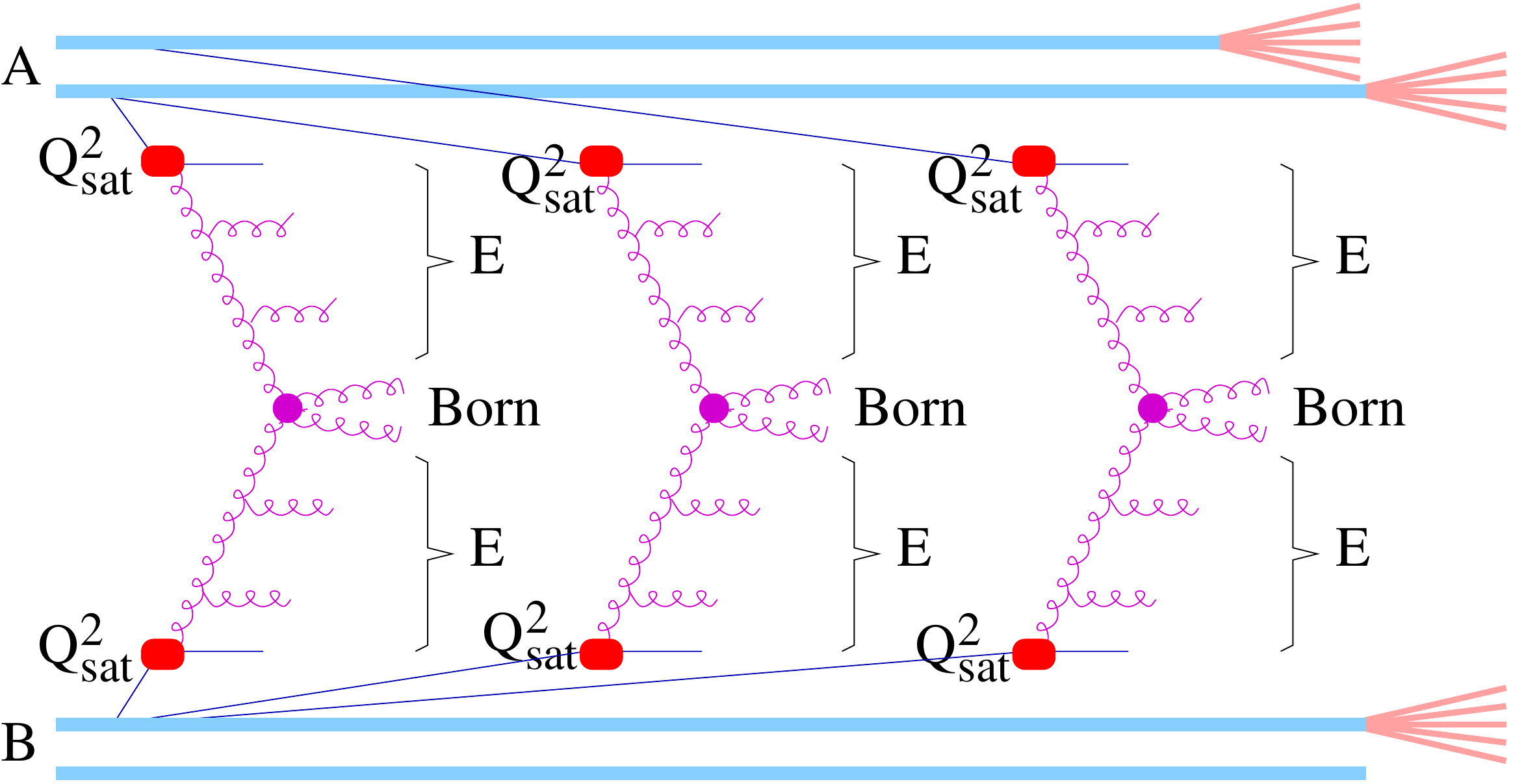}
\caption{Rigorous parallel scattering scenario, for $n=3$ parallel scatterings
for a collision of a nucleus $A$ with a nucleus $B$ , including
non-linear effects via saturation scales. \label{saturation-three-pom-1}}
\end{figure}

Validity of AGK means that we can do the same as models based on factorization
(defining nand using parton distribution fuctions) to study very hard
processes, but this represents only a very small fraction of all possible
applications, and there are very interesting cases outside the applicability
of that approach. A prominent example, one of the highlights of the
past decade in our domain, concerns collective phenomena in small
systems. It has been shown that high-multiplicity pp events show very
similar collective features as earlier observed in heavy ion collisions
\cite{CMS:2010ifv}. High multiplicity means automatically ``multiple
parton scattering''. As discussed earlier, this means that we have
to employ the full parallel scattering machinery developed earlier,
based on S-matrix theory. We cannot use the usual parton distribution
functions (representing the partonic structure of a fast nucleon),
we have to treat the different scatterings (happening in parallel)
individually, for each one we have a parton evolution according to
some evolution function $E$ (representing the partonic structure
of a fast parton), as sketched in Fig. \ref{saturation-three-pom}.
We still have DGLAP evolution, for each of the scatterings (we only show the spacelike (SL) cascade), 
but we introduce saturation scales. But, most importantly, these scales are
not constants, they depend on the number of scatterings, and they
depend as well on $x^{+}$ and $x^{-}$.
An example of a multiple
scattering AA configuration is shown in Fig. \ref{saturation-three-pom-1}.

\subsection{EPOS4 heavy quark issues}

\begin{figure}[h]
\noindent \centering{}(a)\hspace*{3cm}(b)\hspace*{3cm}(c)\hspace*{6cm}$\qquad$\\
 \includegraphics[scale=0.25]{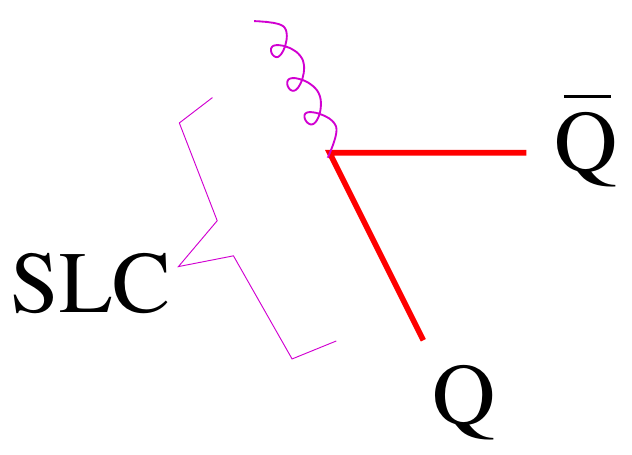}$\qquad$\includegraphics[scale=0.25]{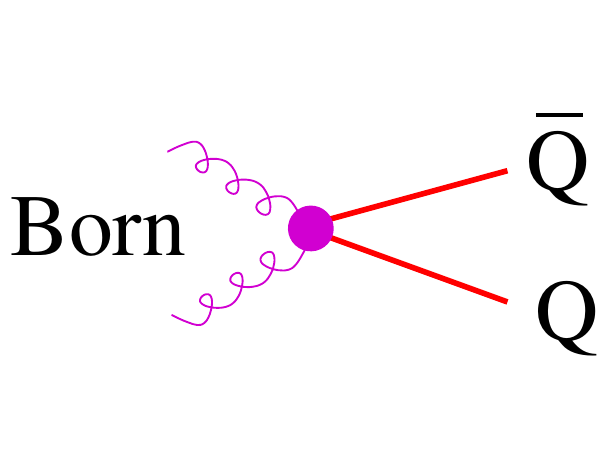}$\qquad$\includegraphics[scale=0.25]{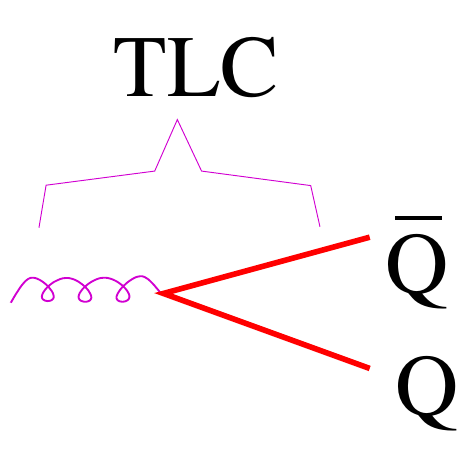}\caption{Different possibilities to create heavy flavor, (a) in the space-like
cascade (SLC), (b) in the Born process, (c) in the time-like cascade
(TLC). \label{charm-3}}
\end{figure}

\begin{figure}[h]
\noindent \centering{}$\qquad$(a)\hspace*{4cm}(b)\hspace*{4cm}$\qquad$\\
 \includegraphics[scale=0.25]{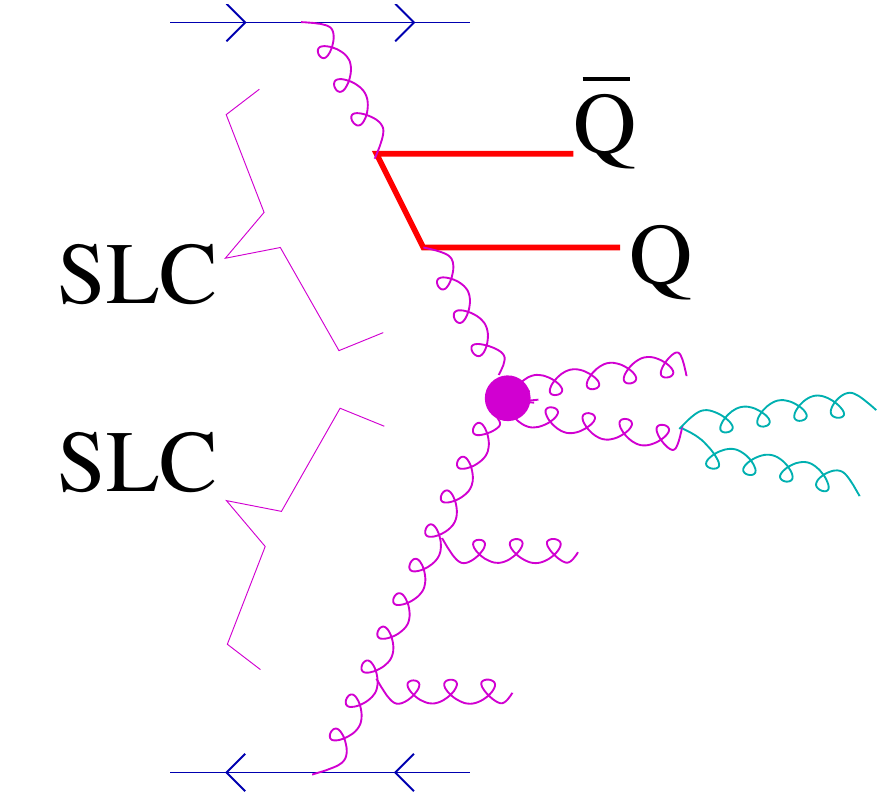}$\qquad$\includegraphics[scale=0.25]{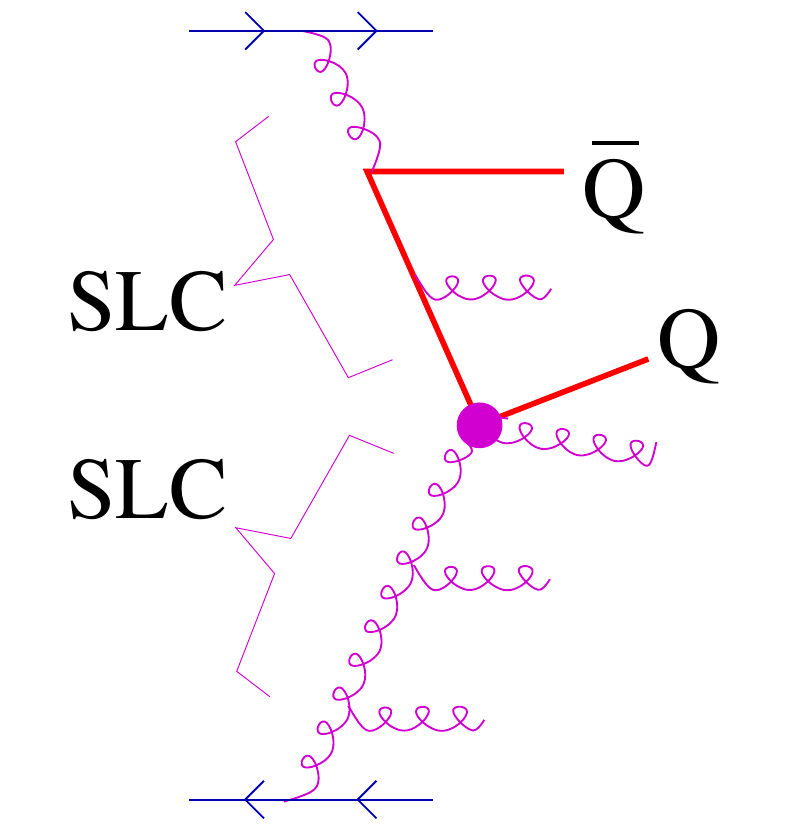}\\
 $\qquad$\hspace*{2cm}(c)\hspace*{7cm}$\qquad$\\
 $\qquad$\includegraphics[scale=0.25]{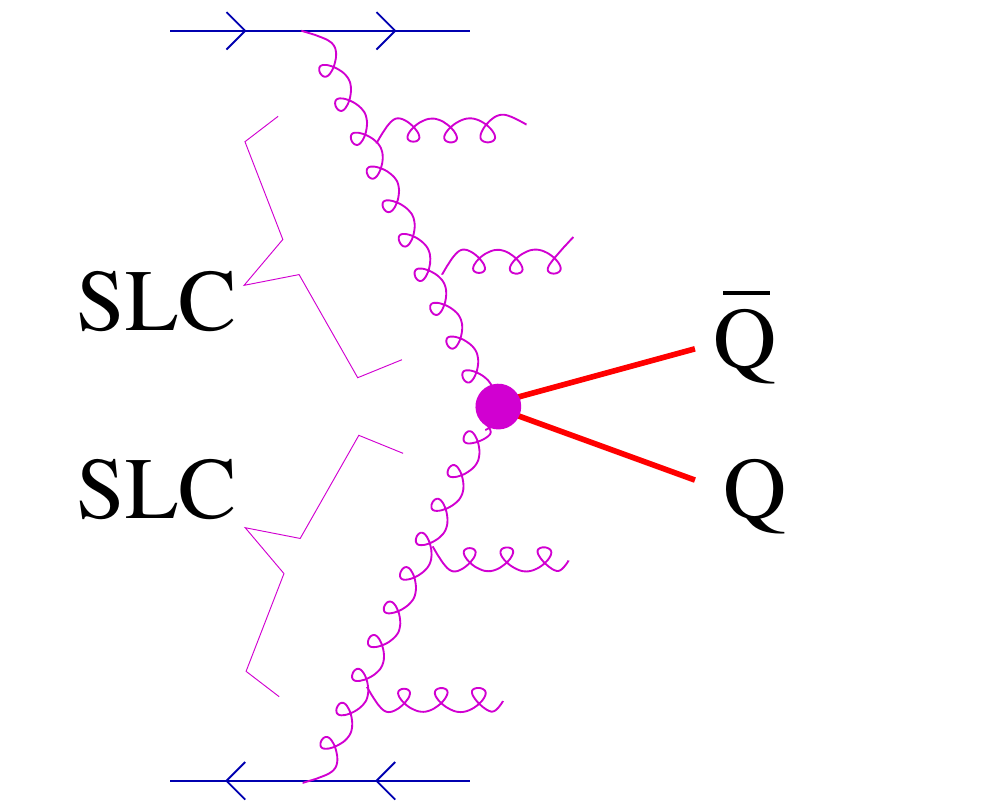}\caption{Heavy flavor production (a,b) in the SL cascade and (c) in the Born
process. The magenta point indicated the Born process. \label{charm-1}}
\end{figure}

\begin{figure}[h]
\noindent \centering{}$\qquad$(a)\hspace*{4cm}(b)\hspace*{4cm}$\qquad$\\
 \includegraphics[scale=0.25]{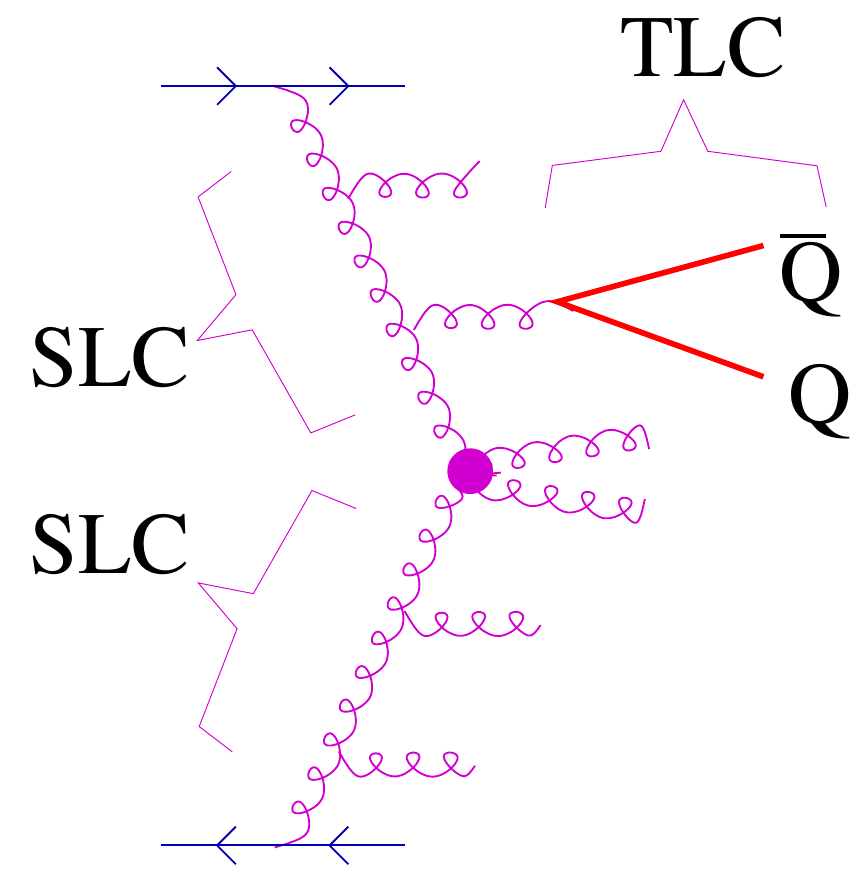}$\qquad$\includegraphics[scale=0.25]{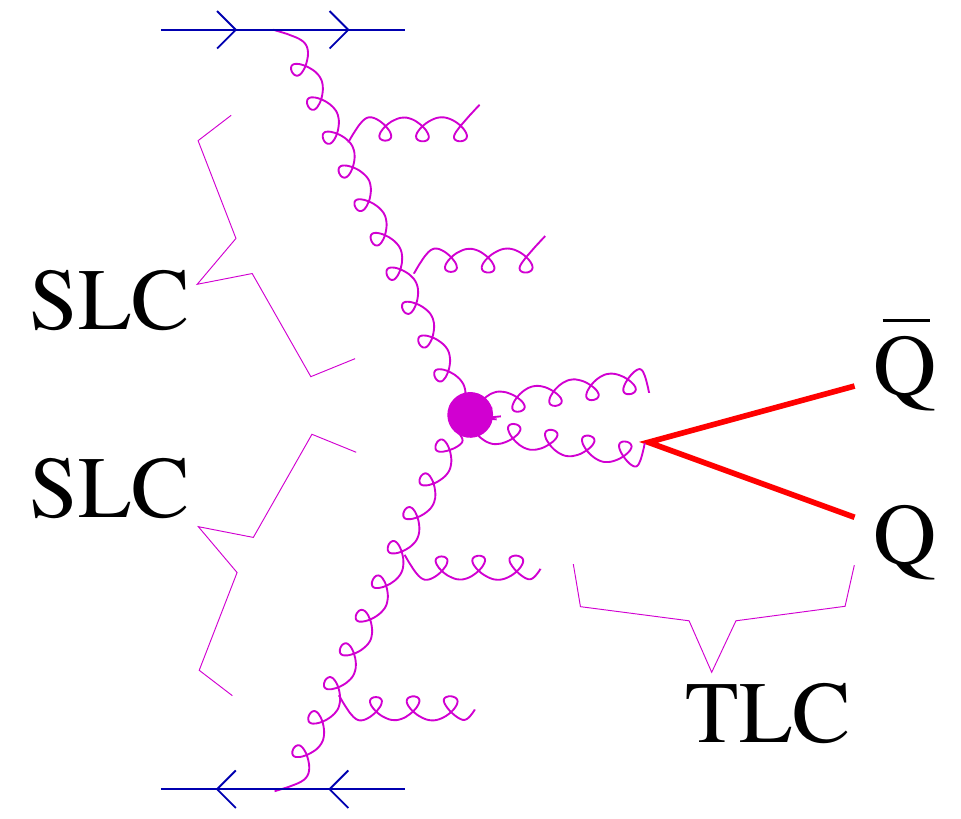}
 \caption{Heavy flavor production in the TL cascade. \label{charm-4-2}}
\end{figure}

At each step in the SL cascade, there is the possibility of quark-antiquark
production, and in the Born process as well. In the following, we
discuss in particular the case of heavy flavor quarks, with the general
notation $Q$ for quarks and $\bar{Q}$ for antiquarks (for details see Ref. \cite{werner:2023-epos4-heavy}). Heavy flavor
may be produced in different ways, as shown in Fig. \ref{charm-3}.
Starting from a gluon, a $Q-\bar{Q}$ pair may be produced in the
SL cascade, as shown in Fig. \ref{charm-3}(a), provided the virtuality
is large enough. The number of allowed flavors is considered to be
depending on the virtuality (variable flavor number scheme). It is
also possible to create a $Q-\bar{Q}$ in the Born process, via $g+g\to Q+\bar{Q}$
or $q+\bar{q}\to Q+\bar{Q}$ (for light flavor quarks $q$), as shown
in Fig. \ref{charm-3}(b), and finally $Q-\bar{Q}$ may be produced
in the TL cascade, via $g\to Q+\bar{Q}$, as shown in Fig. \ref{charm-3}(c).
Let us first consider the $Q-\bar{Q}$ production in the SLC. We may
have the situation as shown in Fig. \ref{charm-1}(a), 
where a heavy flavor parton (here a $\bar{Q}$) is emitted, and the
corresponding antiparticle (here a $Q$) continues the SLC. But before
reaching the Born process, it is emitted, and a gluon continues the
SLC. The two heavy flavor partons have in general low transverse momenta.
Another possibility is shown in Fig. \ref{charm-1}(b), where a heavy
flavor parton produced in the SLC ``survives'' till the Born process,
and the latter has most likely the form $Q+l\to Q+l$, with $l$ being
a light flavor parton. Other than the production during the SLC, heavy
flavor may be produced in the Born process, via $g+g\to Q+\bar{Q}$
or $q+\bar{q}\to Q+\bar{Q}$ (for light flavor quarks $q$), as shown
in Fig. \ref{charm-1}(c). Finally, heavy flavor may be produced during
the time-like cascade, as shown in Fig. \ref{charm-4-2}, 
either initiated from a TL parton in the SL cascade (Fig. \ref{charm-4-2}(a)),
or initiated from an outgoing parton of the Born process (Fig. \ref{charm-4-2}(b)).
In the first case, the transverse momenta are in general small.

\begin{figure}[h]
\noindent \centering{}\includegraphics[scale=0.25]{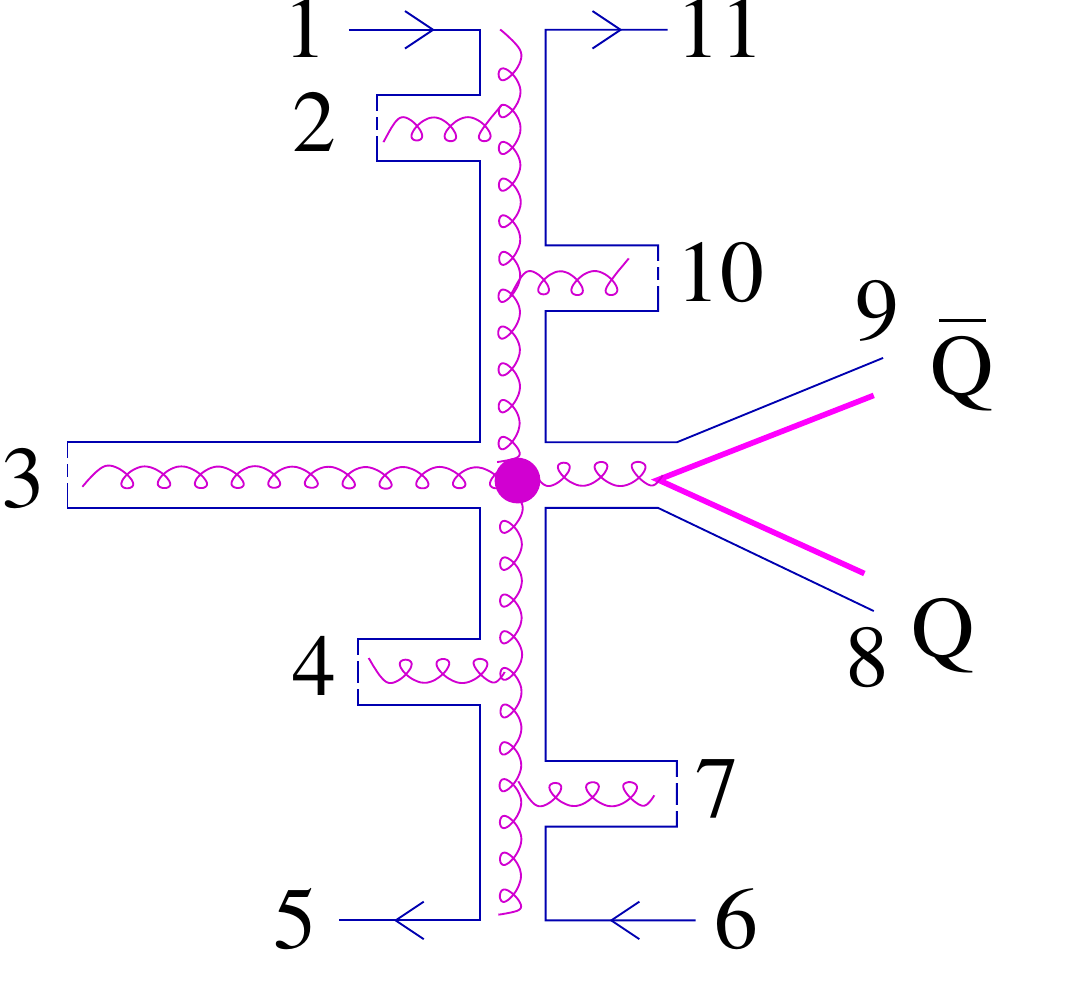}
\caption{A possible color flow diagram corresponding to the graph of Fig. \ref{charm-4-2}(b)
\label{charm-4-1}}
\end{figure}

The next step will be, for a given Feynman diagram, to construct the
color flow diagram. Let us
take the graph of Fig. \ref{charm-4-2}(b), i.e. heavy flavor production
during the TLC of an outgoing Born parton. As usual, the gluons are
emitted to either side with equal probability, so a possible color
flow diagram is the one shown in Fig. \ref{charm-4-1}. 
We identify three chains of partons: 1-2-3-4-5, 6-7-8, and 9-10-11.
The initial TL partons (the horizontal blue lines with arrows) or
most likely quarks and antiquarks (in any case $3$ and $\bar{3}$
color representations). Let us assume that 3 is a quark, and 6 an
antiquark (light flavor, both), then the two chains containing heavy
flavor are of the form $\bar{Q}-g-q$ and $\bar{q}-g-Q$, in both
cases, the heavy flavor partons are ``end partons'' in the chains.

\begin{figure}[h]
\noindent \centering{}\includegraphics[scale=0.25]{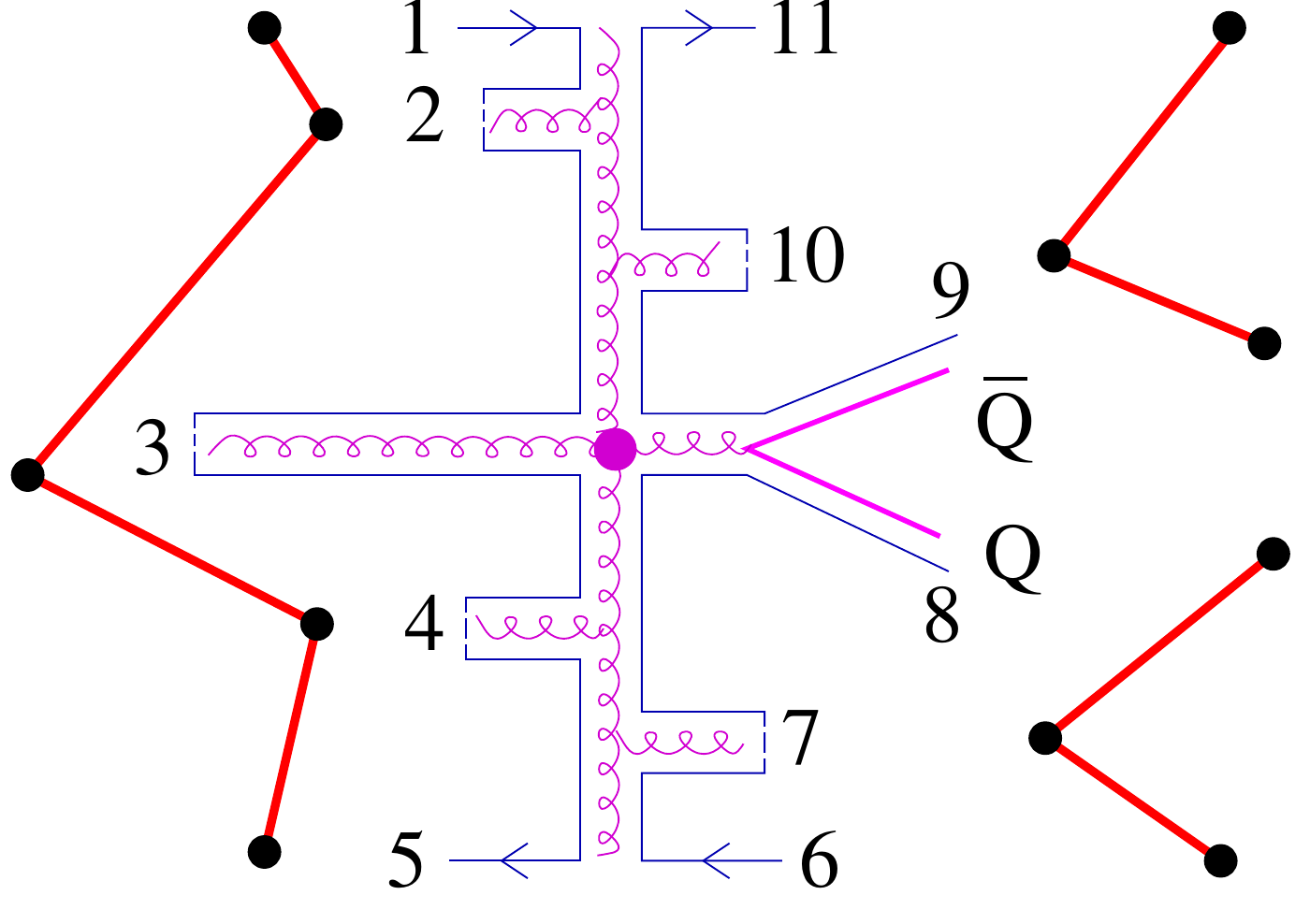}
\caption{The chains 1-2-3-4-5, 6-7-8, and 9-10-11 are mapped to kinky strings
(red lines). The black points indicate the kinks, which carry the
parton momenta. 
\label{charm-4-1-1}}
\end{figure}

These chains of partons are finally mapped (in a unique fashion) to
kinky strings, where each parton corresponds to a kink, as shown in
Fig. \ref{charm-4-1-1}.
The general mapping procedure (chains of partons to kinky strings)
as well as the string decay procedures (producing so-called ``prehadrons'') are described in detail in
\cite{Drescher:2000ha}.

\subsection{EPOS4 core-corona method and fluid evolution }

The above-mentioned parallel scattering happens at zero time. After that, we obtain a more or
less important number of prehadrons. We employ a core-corona procedure
\cite{Werner:2007bf,Werner:2010aa,Werner:2013tya,werner:2023-epos4-micro}, where the prehadrons,
considered at a given proper time $\tau_{0}$, are separated into
``core'' and ``corona'' prehadrons, depending on the energy loss
of each prehadron when traversing the ``matter'' composed of all
the others. Corona prehadrons (per definition) can escape, whereas
core prehadrons lose all their energy and constitute what we call
``core'', which acts as an initial condition for a hydrodynamic
evolution~\cite{Werner:2013tya,Karpenko_2014}.
\begin{figure}[h]
\begin{centering}
\includegraphics[bb=0bp 0bp 567bp 520bp,clip,scale=0.36]{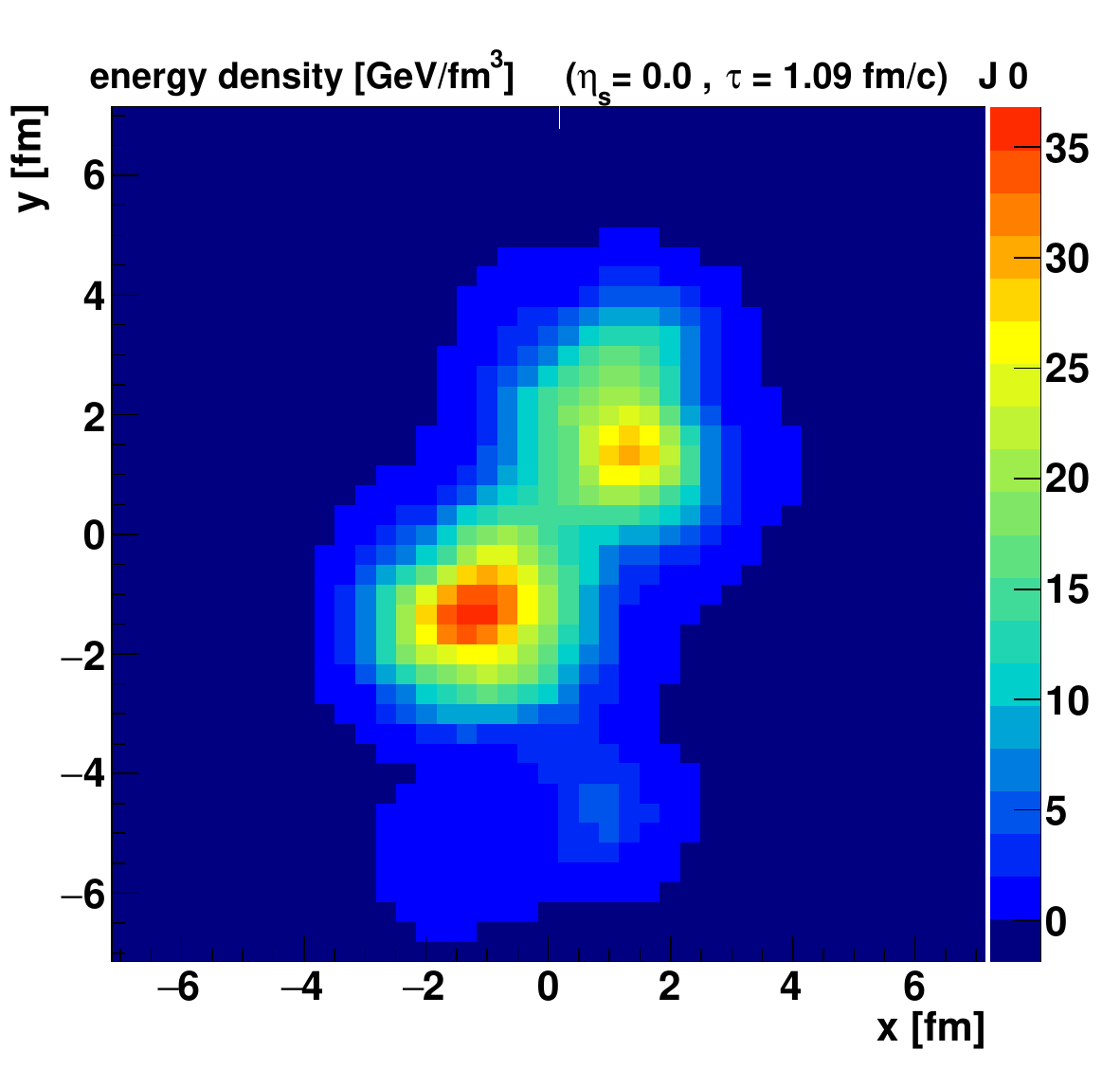}
\par\end{centering}
\begin{centering}
\includegraphics[bb=0bp 0bp 567bp 520bp,clip,scale=0.36]{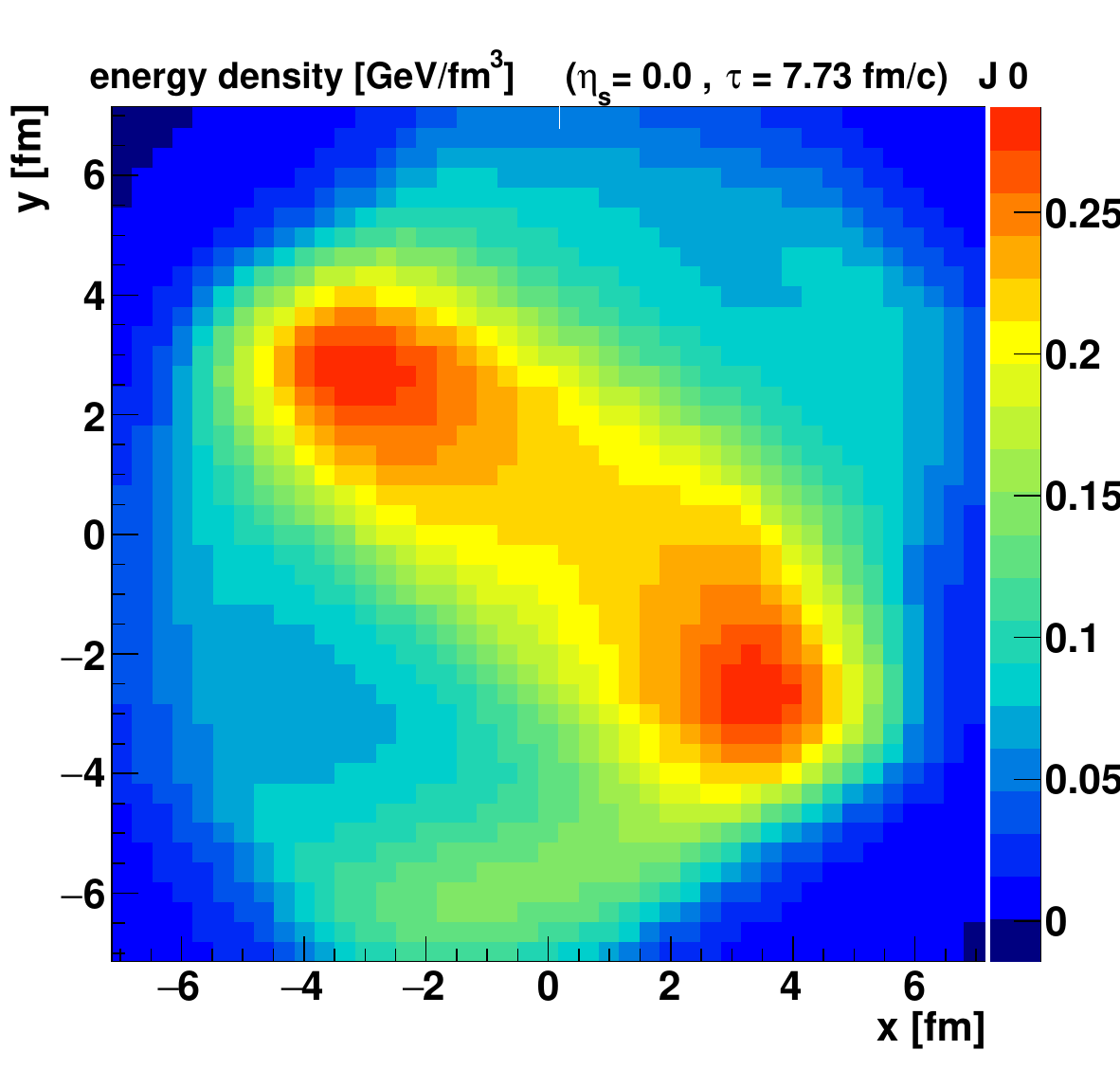}\\
\par\end{centering}
\centering{} \caption{ Energy density in the transverse plane ($x,y)$ for a Pb-Pb collision at $\sqrt{s_{\rm NN}}=5.02~\rm TeV$ with an impact paremeter of 10.4 fm.
The upper plot represents the start time $\tau_{0}$ (of the hydro evolution), and the lower plot a later
time $\tau_{1}$, close to final freeze-out.
\label{Energy-density-6-Pomerons}}
\end{figure}
The evolution of the core ends whenever the energy density falls below
some critical value $\epsilon_{\mathrm{FO}}$, which marks the point
where the fluid ``decays'' into hadrons. It is not a switch from
fluid to particles, it is a sudden decay, called ``hadronization''.
Let us consider a (randomly chosen, but typical)  5.02 ATeV lead-lead  scattering. In Fig. \ref{Energy-density-6-Pomerons}, we plot the
energy density in the transverse plane ($x,y)$. We consider two snapshots,
namely at the start time of the hydro evolution $\tau_{0}$ (upper plot) and a later time $\tau_{1}$ close to final freeze-out
(lower plot). The initial distribution has an elongated shape (just
by accident, due to the random positions of interacting partons).
One can clearly see that the final distributions are as well elongated,
but perpendicular to the initial ones, as expected in a hydrodynamical
expansion. More examples can be found in \cite{werner:2023-epos4-micro}.

In EPOS4, as discussed in detail in \cite{werner:2023-epos4-micro},
we developed a new procedure of energy-momentum flow through the ``freeze-out
(FO) hypersurface'' defined by $\epsilon_{\mathrm{FO}}$, which allows
defining an effective invariant mass,  decaying according to microcanonical
phase space into hadrons, which are then Lorentz boosted according
to the flow velocities computed at the FO hypersurface. We also developed
new and very efficient methods for the microcanonical procedure \cite{werner:2023-epos4-micro}.
Also in the full scheme, including primary and secondary interactions,
energy-momentum and flavors are conserved.

\section{initial charm quark momentum distribution}
\label{sec.initial}
EPOS2 and EPOS3 provided only the interaction points at which charm and anticharm quarks have been produced. Their transverse momentum had been chosen to reproduce the FONLL calculations~\cite{Cacciari:1998it,Cacciari:2005rk} and energy and momentum has not been conserved at the interaction points. Correlations between the heavy quark and antiquark have not been considered.
 
\begin{figure}[!htb]
\includegraphics[width=0.4\textwidth]{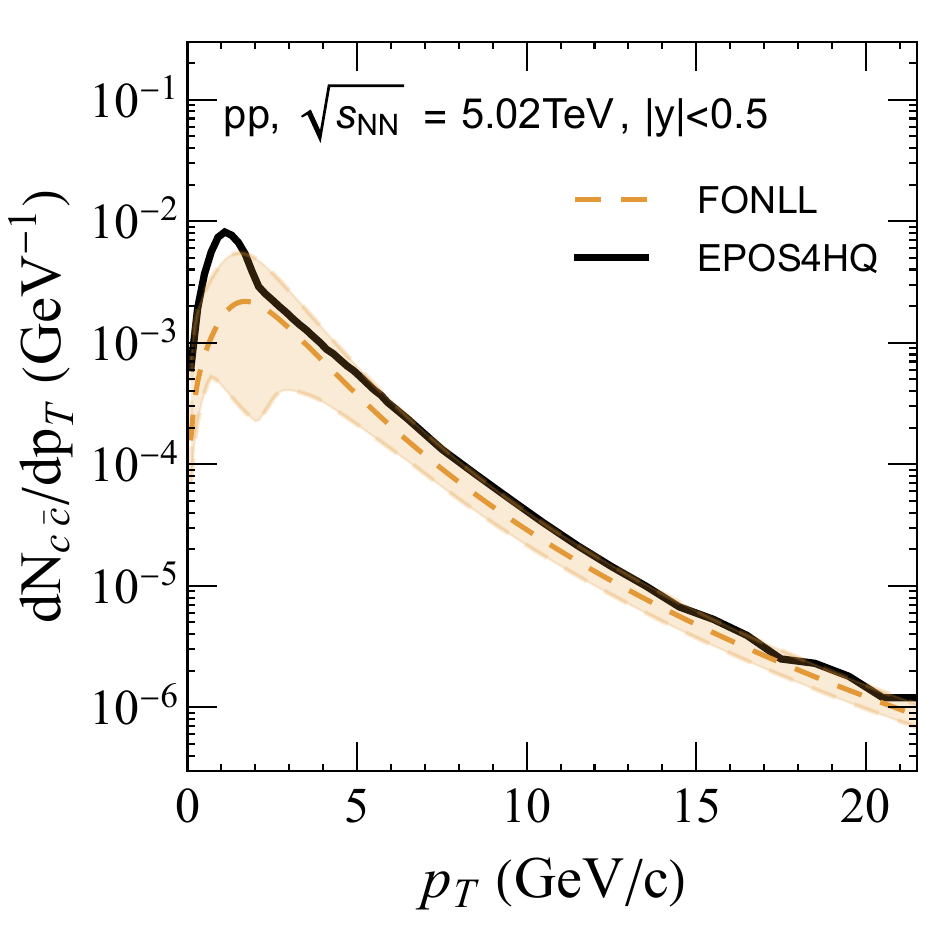}\\
\includegraphics[width=0.4\textwidth]{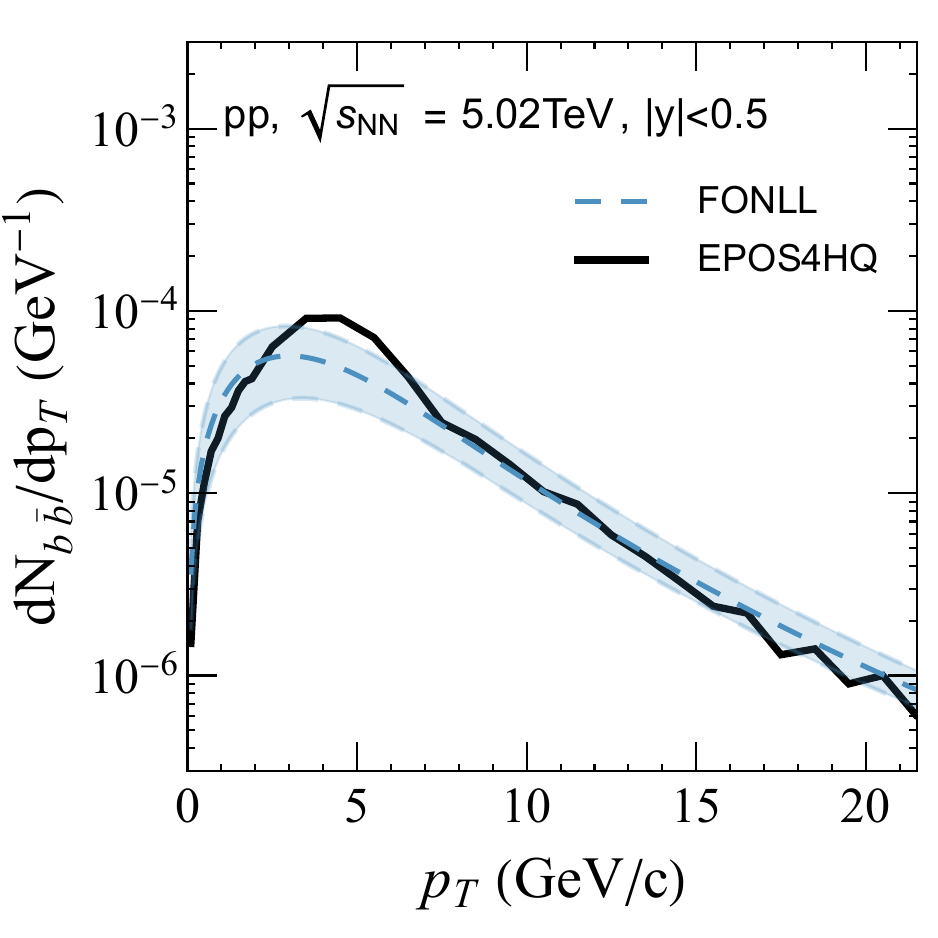}
\caption{The initial transverse momentum distribution of charm (upper) and bottom (lower) quarks for pp collisions at 5.02TeV. The solid black line is the EPOS4HQ calculation, while the orange (blue) dashed line presents the FONLL~\cite{Cacciari:1998it} result. The uncertainty of the FONLL result is given by the orange (blue) shaded area.}
\label{fig_ccbar}
\end{figure}

\begin{figure}[!htb]
\includegraphics[width=0.4\textwidth]{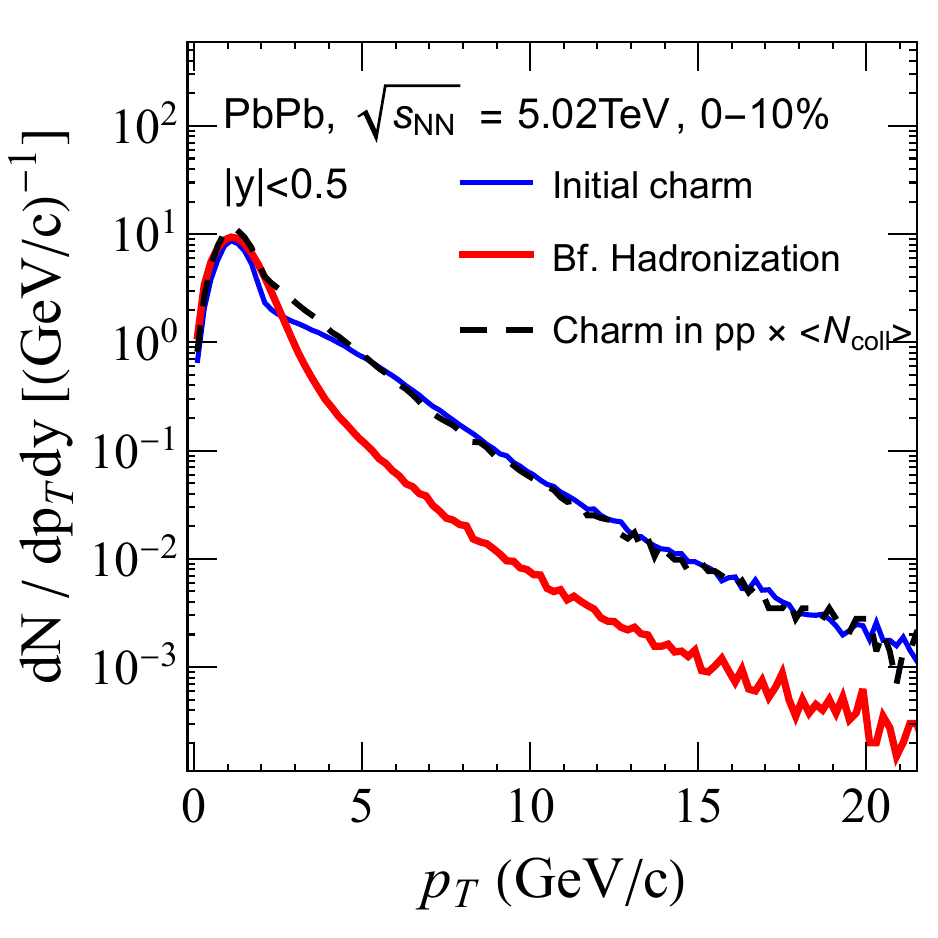}
\caption{The transverse momentum distribution of charm quarks at creation and before hadronization for central  PbPb collisions at $\sqrt{s_{\rm NN}}=$ 5.02 TeV . The transverse momentum distribution observed in pp collisions, multiplied by the number of initial binary collisions, is shown as a black dashed line.}
\label{fig_charm_PbPb}
\end{figure}

EPOS4, on the contrary, provides the correlation between the  momenta of the charm and the anticharm quark, which are created at the same vertex. These correlations are difficult to compare with other approaches. Only PYTHIA provides this correlation as well. In the standard version of PYTHIA, however,  only the leading order cross section is calculated and therefore the opening angle between $c$ and $\bar c$ is 180°.  Including initial and final state interaction this distribution is modified but a one to one correspondence with the diagrams included in EPOS4 is not possible. The initial single (anti)charm quark  $p_T$ distribution, given by EPOS4HQ for pp collision can be  compared with the fixed order next to leading log (FONLL)  calculations \cite{Cacciari:1998it,Cacciari:2005rk}, the most advanced pQCD approach for describing single charm quark distributions. In Fig.~\ref{fig_ccbar} we display the FONLL and EPOS4 single charm and bottom quark distribution  for 5.02 TeV pp collisions. We see that the EPOS4 distribution is at the upper limit of the error bars of the FONLL calculation for low $p_T$ and consistent with FONLL for high $p_T$. This means that EPOS4 gives a larger charm production cross section, which is consistent with the experimental data obtained recently~\cite{ALICE:2023sgl}. When produced in heavy-ion collisions, shadowing, saturation effects and the Cronin effect modify the initial (anti)charm quark distribution. In Fig.~\ref{fig_charm_PbPb} we compare for PbPb at $\sqrt{s_{\rm NN}}$ = 5.02 TeV the extrapolated pp $p_T$ distribution (black dashed line) with the PbPb $p_T$ distribution, which takes the above mention cold nuclear matter effects into account (blue line). At large $p_T$ both distribution are almost identical but at low $p_T$ one sees differences.

\section{EPOS4HQ: Heavy quark energy loss in hot medium}
\label{sec.epos4hq}
In contradistinction to EPOS4, in EPOS4HQ heavy quarks interact with the partons of the QGP, formed by the light partons and gluons from the core. We include in this study both, elastic~\cite{Gossiaux:2008jv} and radiative ~\cite{Aichelin:2013mra} collisions. To determine where the interaction takes place, we calculate the
interaction rate and move the heavy quark to the interaction point, select whether the QGP parton is a gluon or a quark and draw the momentum of the scattering QGP parton from its corresponding thermal distribution. The thermal distribution is determined by the local temperature and mean velocity at the position at which the collision takes place. The scattering cross sections of the heavy quark with gluons and light quarks are calculated by pQCD matrix elements with a running coupling constant.

The pQCD elastic scattering cross section diverges for a small momentum transfer in the $t$ and $u$ channels. These infrared divergences are healed by the Debye screening mass $m_D(T)$ of gluons in the hot medium, which is calculated  in the hard thermal loop (HTL) approach.  It serves as a regulator of the propagator of the exchanged gluon.  Scattering at high momentum transfer is, on the contrary,  described by a free gluon propagator for massless gluons. A smooth transition of the energy loss between both regimes can be assured by an effective Debye mass $m_{\rm eff}=\kappa m_D(T)$ in the gluon propagator, with $\kappa=0.2$~\cite{Gossiaux:2008jv}.

The pQCD inelastic scattering cross section has been calculated in~\cite{Aichelin:2013mra}.
This cross section contains 5 matrix elements for gluon emission from the heavy quark  and the light quark and gluon, respectively.  Also for the inelastic cross section the momentum of the plasma particle is chosen by a Monte Carlo approach from the local thermal distribution. 
As in the elastic cross section, the gluon propagator is regulated by $m_{\rm eff}=\kappa m_D(T)$.  For the gluon emission vertex a constant $\alpha_S = 0.3$ is used. The emitted gluon is considered as massless. The different limits of the pQCD cross section calculations as well as more details of the approach have been discussed in Ref.~\cite{Aichelin:2013mra} .

\begin{figure}[!htb]
\includegraphics[width=0.4\textwidth]{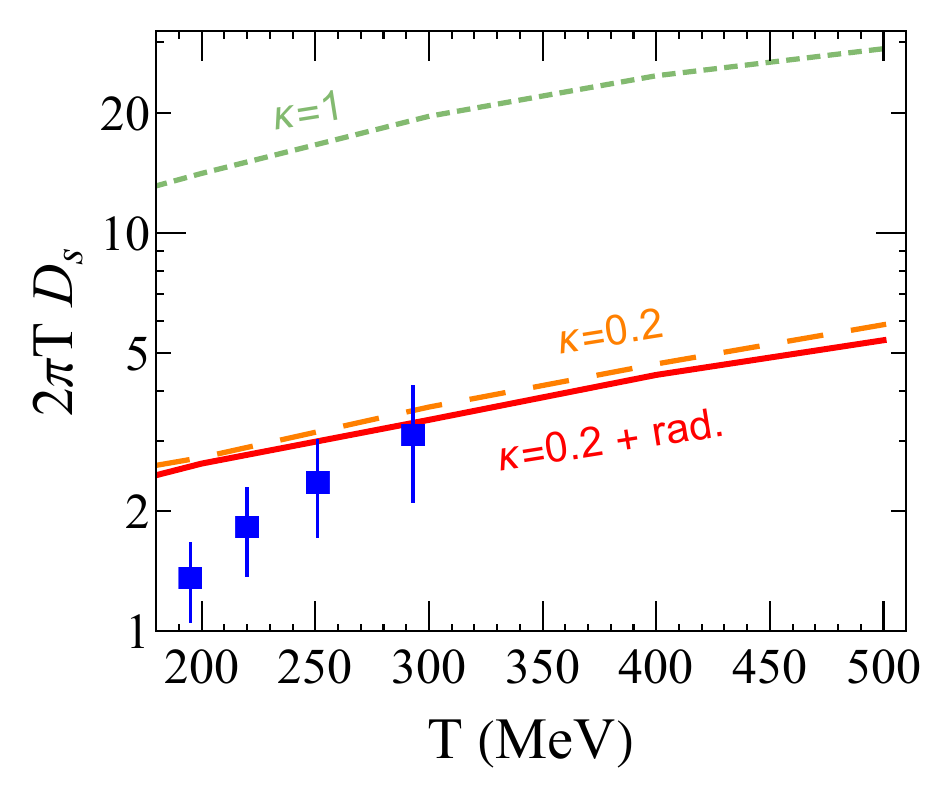}
\caption{$2\pi T \mathcal{D}_s$ obtained in our approach in comparison with lattice data from \cite{Altenkort:2023eav}. The plain red line corresponds to our collisional+radiative model ($K=1$), while the orange long-dashed line represents the contribution from the sole collisional part; the dashed curves illustrates the equivalent calculation if the IR regulator is taken as the Debye mass ($\kappa=1$).}
\label{fig_lat}
\end{figure}

The validity of this approach can be confirmed by comparing its spatial diffusion coefficient $\mathcal{D}_s$ \cite{Berrehrah:2014tva} with lattice calculations. Fig.~\ref{fig_lat} shows $2\pi T \mathcal{D}_s$ of our model, in comparison with recent unquenched lattice data~\cite{Altenkort:2023eav}. The former lattice data for $2\pi T \mathcal{D}_s$ were calculated in a quenched approximation, so we do not include them here. We see that both coefficients agree well for $\kappa$ =0.2 whereas there are large differences for the choice $\kappa$ = 1. Inelastic collisions have only a negligible influence on
$\mathcal{D}_s$. 

Both, the elastic as well as the inelastic collisions, have been already employed in the EPOS2 and EPOS3 framework to described heavy meson data in heavy-ion collisions at LHC energies ~\cite{Nahrgang:2013saa,Nahrgang:2014vza}.  
We use this theoretical framework without modification also in this new EPOS4HQ version. In this paper the K-factor for elastic as well as for inelastic collisions, which has been varied in the past Refs.~\cite{Nahrgang:2013saa,Nahrgang:2014vza} in calculations in which the HQ part was coupled to EPOS2 or EPOS3, is equal one, so the calculated pQCD cross sections are not modified by an overall factor. 

Fig.~\ref{fig_charm_PbPb} displays the change of the momentum distribution of heavy quarks caused by the interactions with the QGP for central PbPb collisions at $\sqrt{s_{\rm NN}}=5.02$ TeV. As said, the blue line is the initial transverse momentum distribution. The $p_T$ distribution of the heavy quark just before they hadronize is represented by the red line. At low $p_T$ there is little difference between the curves because the initial $p_T$ of the heavy quarks is close to the value one expects if the heavy quarks come to a thermal equilibrium with the light quarks, so the momentum change is small. For large momenta, on the contrary, we observe a $p_T$ shift to lower $p_T$ values of the order of 5 GeV. 

\section{Hadronization scheme}
\label{sec.hadronization}
When the  QGP has expanded to the critical value of the  local temperature or energy density  ($T_{\rm FO}=167 \rm MeV$ or $\epsilon_{\rm FO}=0.57 \rm GeV/fm^3$), the system hadronizes and the heavy quark converts into a heavy flavor hadron. In the EPOS4HQ framework, there are two  ways in which the  heavy quark can hadronize, either by fragmentation or by coalescence.
In the coalescence process, the heavy quarks coalesce with light quarks from the hypersurface at which they are localized.  The fluid at the hypersurface has an average velocity. The momentum of the light quark, with which the parton coalesce, is selected randomly from the thermal distribution with the fluid temperature $T_{FO}$ and the average fluid velocity at the hypersurface.

In this study, we neglect the coordinate information of quarks and require only that the hadronization hypersurface is same for the heavy and light quarks. In the numerical simulation the coalescence formula is applied in the center-of-mass frame of heavy and light quarks.

The momentum distribution of HF hadrons, produced via the coalescence process, can be calculated in the thermal rest system by:
\begin{eqnarray}
{dN\over d^3{\bf P}}&=&g_H\sum_{N_Q}\int \prod_{i=1}^k{d^3p_i\over (2\pi)^3} f({\bf p}_i)\nonumber\\
&\times&W_H({\bf p}_1,..,{\bf p}_i)\, \delta^{(3)}\left({\bf P}-\sum_{i=1}^N{\bf p}_i\right),
\label{eq.coal}
\end{eqnarray}  
where $g_H$ is the degeneracy factor of color and spin. ${\bf P}$ and ${\bf p}_i$ are the momenta of heavy flavor hadrons and the constituent quarks, respectively. The delta function conserves momentum. The integration is carried at the point where the heavy quark crosses the hadronization hypersurface, which is given by EPOS4. The summation is performed over all heavy quarks in the system. $f({\bf p}_i)$ is the momentum space distribution of the constituent in the heavy hadron at the moment of hadronization with $k=2$ for mesons and 3 for baryons.
$W_H({\bf p}_1,..,{\bf p}_k)$ is the Wigner density of a given heavy hadron $H$, which can be constructed from the heavy hadron wave function, the solution of the Schrödinger equation, and which is approximated here by a three-dimensional harmonic oscillator state with the same root mean square radius. After integration over the coordinate space, the Wigner density of the mesonic  ground state can be expressed as
\begin{eqnarray}
W(p_r)&=&{(2\sqrt{\pi}\sigma)^3}e^{-\sigma^2{p_r}^2}.
\end{eqnarray} 
 $p_r$ is the relative momentum between the two constituent quarks in the center-of-mass (CM) frame, $
 p_r=|E_2{\bf p}_1-E_1{\bf p}_2| /(E_1+E_2)$, $E_1 ({\bf p}_1)$ and $E_2({\bf p}_2)$ are the energies (momenta) of the quark and antiquark in the heavy hadron CM frame, respectively. Baryons are treated as two two-body systems (baryons are produced by recombining two particles first and then by using their center of mass to recombine this diquark with the third quark).

\begin{figure}[!htb]
\includegraphics[width=0.4\textwidth]{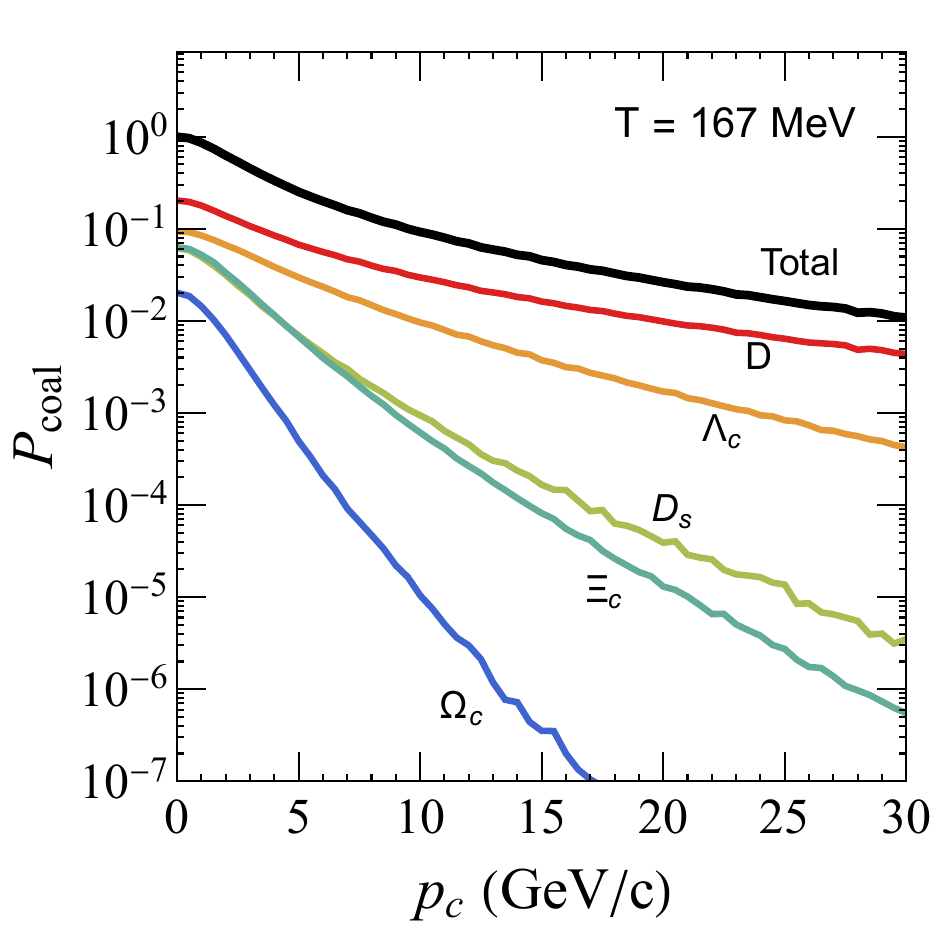}\\
\includegraphics[width=0.4\textwidth]{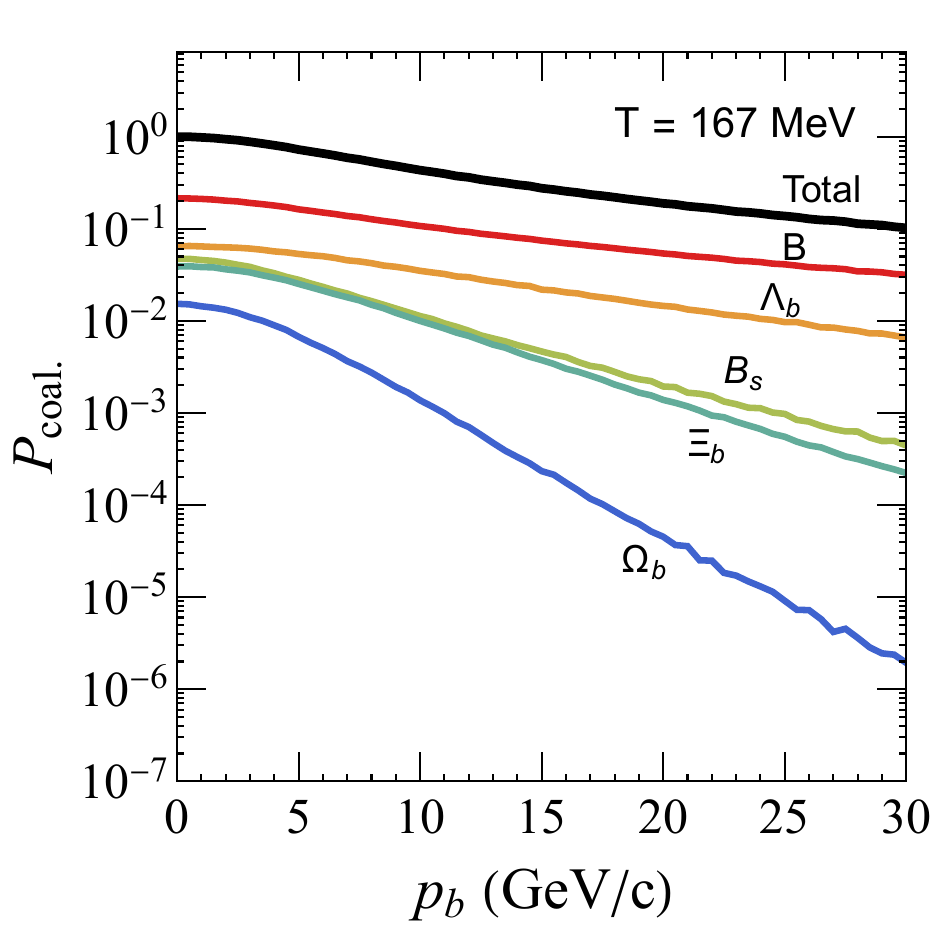}
\caption{The momentum dependent coalescence probabilities of charm (upper) and bottom (lower) quark in a hot and static medium with temperature $T=167\rm MeV$.}
\label{fig_coal_prob}
\end{figure}
The light quarks (antiquarks) are assumed to be thermalized.  In the rest system of the hypersurface their distribution is given by,
\begin{eqnarray}
f_q({\bf p}_q)={1\over (2\pi)^3}{g\over e^{E_q/T}+1},
\label{eq.light}
\end{eqnarray}  
where $g=6$ is the statistical factor. $E_q=\sqrt{{\bf p}_q^2+m_q^2}$ is the energy of the light quark. The quark masses are $m_{u/d}=0.1\rm GeV$, $m_{s}=0.3\rm GeV$.

The heavy quark coalescence probability $P_{\rm coal}$ in a static hot medium with a temperature $T_{\rm FO}=167 \rm MeV$ for the ground states of the open heavy flavor states, e.g. $D^0$, $D^+$, $D_s$, $\Lambda_c$, $\Xi_c$, and $\Omega_c$ for charm, $B^0$, $B^-$, $B_s$, $\Lambda_b$, $\Xi_b$, and $\Omega_b$ for bottom, is obtained by integrating Eq.~\eqref{eq.coal} with a delta distribution $\delta^{3}({\bf p}_1-{\bf p}_Q)$ for the momentum of the heavy quark and with a delta distribution $\delta^{3}({\bf p}_2-{\bf p}_q)$ for the momentum of the quark. $p_q$ is sampled from the thermal distribution, Eq.~\eqref{eq.light}. We neglect here the rarely produced heavy flavor hadrons, such as multi-charmed (bottomed) baryons, $B_c$, and quarkonium.
For heavy flavor mesons, the averaged square radius can be expressed as $\langle r^2\rangle={3\over 2}{m_Q^2+m_q^2\over (m_Q+m_q)^2}\sigma^2$ with charm quark mass $m_{c}=1.5\rm GeV$ and $m_{b}=4.5\rm GeV$ for bottom quarks. The root-mean-square radius of the ground state charmed meson has been calculated by the two-body Dirac equation~\cite{Zhao:2018jlw}. It gives $\sqrt{\langle r^2\rangle}=0.85\rm fm$, for in-medium $D^0$. So, the corresponding width $\sigma=3.725\rm GeV^{-1}$ for $D^0$. In the absence of theoretical studies of the in-medium radius of charmed baryon, we take the same width $\sigma$ for any two-quark systems in charmed baryons. 
Because the reduced masses are very similar, the averaged radius of bottom mesons is comparable to that of charmed mesons. So, we take the same width for bottom mesons and any two-quark systems in bottom baryons. 

The coalescence probability of excited states, which can strongly decay into the ground states, is estimated via the statistic model. There the hadron density at the temperature $T_{\rm FO}$ is given by~\cite{Andronic:2007zu},
\begin{eqnarray}
n_i={g_i\over2\pi^2}T_{\rm FO}m_i^2K_2\left(m_i\over T_{\rm FO} \right),
\end{eqnarray}
where $g_i$ is the spin isospin degeneracy. $m_i$ is the mass of the hadron. $K_2$ is the second-order Bessel function. In our study, we consider almost all possible excited states, also the missing baryons, which are predicted by the quark model~\cite{Ebert:2011kk} and lattice QCD~\cite{Bazavov:2014yba,Padmanath:2014lvr}, as shown in Table~\ref{tab1}. For each 
ground state hadron $D$, $D_s$, $\Lambda_c$, $\Xi_c$, and $\Omega_c$  we calculate the density of their excited states $m$ and define $R^m=n_{\rm excited}^m/n_{\rm ground}$. This ratio is momentum-independent.  Finally we sum up $R=\sum R^m$ and multiply the ground state momentum-dependent coalescence probability by $1+R$ to obtain its effective momentum distribution. The sum of the effective momentum distributions for all hadrons gives the
total coalescence probability $P_{\rm coal}(p)$. Same as for the bottom sector.

\begin{table}[!bt]
\renewcommand\arraystretch{1.5}
\caption{Number of excited states considered in EPOS4HQ ($N_{\rm excited}$) and their relative contributions to the ground states, $R_i$.}
\label{tab1}
\setlength{\tabcolsep}{2.5mm}
\begin{tabular}{ccccccc}
	\toprule[1pt]\toprule[1pt] 
	& $D^0$ & $D^+$ & $D_s$ & $\Lambda_c$ & $\Xi_c$ & $\Omega_c$ \\
	$N_{\rm excited}$ & 9 & 8 & 5 & 91 & 91 & 53\\
 	$R$ & 2.91 & 0.96 & 2.06 & 7.63 & 3.97 & 4.22\\
	\midrule[0.7pt]
	& $B^0$ & $B^-$ & $B_s$ & $\Lambda_b$ & $\Xi_b$ & $\Omega_b$\\
    $N_{\rm excited}$ & 6 & 6 & 2 & 91 & 91 & 53\\
 	$R$ & 2.95 & 2.95 & 2.26 & 7.67 & 3.92 & 5.99\\
	\bottomrule[1pt]\bottomrule[1pt]
\end{tabular}
\end{table}

The probability that  a charm  (bottom) quark with a given momentum ${\bf p}_c$(${\bf p}_b$) forms a specific hadron by coalescence  is shown in Fig.~\ref{fig_coal_prob}. Heavy quarks, which do not hadronize via  coalescence, will fragment into a heavy-flavor hadron. The fragmentation probability is therefore $1-P_{\rm coal}$. In EPOS4HQ, we use the heavy quark effective theory-based fragmentation function~\cite{Braaten:1994bz,Cacciari:2005rk}. The fragmentation ratios to various charmed hadrons are taken as the $e^+e^-$ collisions~\cite{Lisovyi:2015uqa} and are shown in Table~\ref{tab2}. 
\begin{table}[!bt]
\renewcommand\arraystretch{1.5}
\caption{Fragmentation ratio of charm (bottom) quark to charmed (bottomed) hadrons in percent.}
\label{tab2}
\setlength{\tabcolsep}{2.5mm}
\begin{tabular}{cccccc}
	\toprule[1pt]\toprule[1pt] 
	$D^0$ & $D^+$ & $D_s$ & $\Lambda_c$ & $\Xi_c$ & $\Omega_c$ \\
	60.8\% & 24.0\% & 8.0\% & 6.0\% &
    1.0\% & 0.2\%\\
	\midrule[0.7pt]
	$B^0$ & $B^-$ & $B_s$ & $\Lambda_b$ & $\Xi_b$ & $\Omega_b$\\
    42.4\% & 42.4\% & 8.0\% & 6.0\% & 
    1.0\% & 0.2\%\\
	\bottomrule[1pt]\bottomrule[1pt]
\end{tabular}
\end{table}
After the hadronization, all charmed hadrons and light hadrons evolve together in the hadronic phase, which is controlled by the UrQMD~\cite{Bass:1998ca}. Bottom hadrons do not interact in the hadronic phase.

\section{Results}
\label{sec.results}
In this section, we present the EPOS4HQ results for open heavy flavour hadrons. We are as exhaustive as possible and show the results for all published experimental data on transverse momentum and elliptic flow. We present as well yield ratios like $R_{AA}$ and interpret these results with help of additional information, which the EPOS4HQ approach provides.

\subsection{Transverse momentum spectra}

\begin{figure}[!htb]
\includegraphics[width=0.46\textwidth]{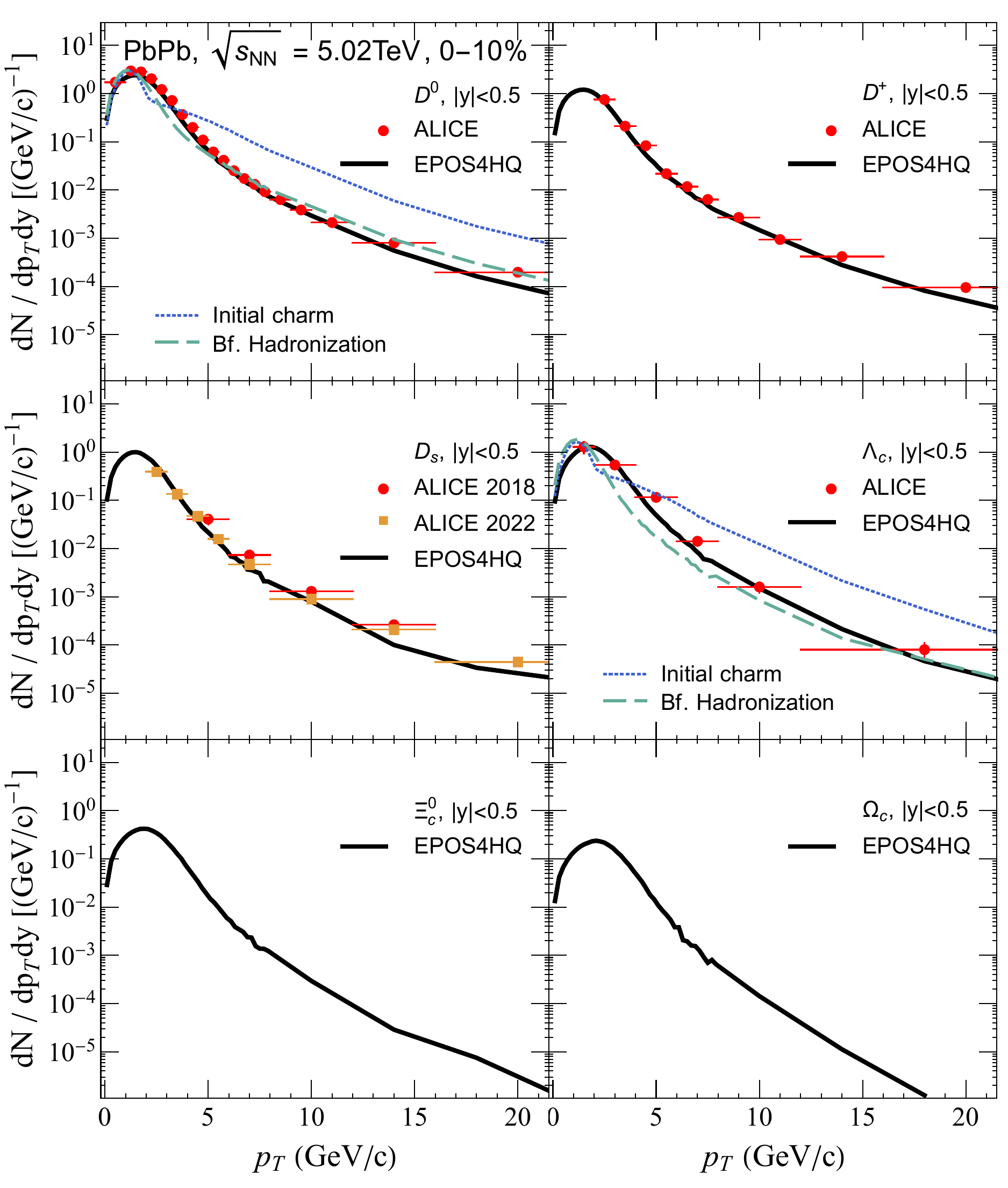}
\caption{$p_T$ spectra of charmed hadrons $D^0$, $D^+$, $D_s$, $\Lambda_c$, $\Xi_c^0$, and $\Omega_c$ in the 0-10\% centrality class of PbPb collisions at $\sqrt{s_{\rm NN}}=5.02\rm TeV$. The experimental data of $D$ and $D^+$~\cite{ALICE:2021rxa}, $D_s$~\cite{ALICE:2018lyv,ALICE:2021kfc}, $\Lambda_c$~\cite{ALICE:2021bib} are from the ALICE. The thick black line is the EPOS4HQ result, the dotted blue and dashed green lines are the $p_T$ distribution of charm quarks at production and before hadronization, respectively, which finally are part of the $D^0$ ($\Lambda_c$).}
\label{fig.spect_PbPb010}
\end{figure}

\subsubsection{charm hadrons}
\begin{figure}[!htb]
\includegraphics[width=0.46\textwidth]{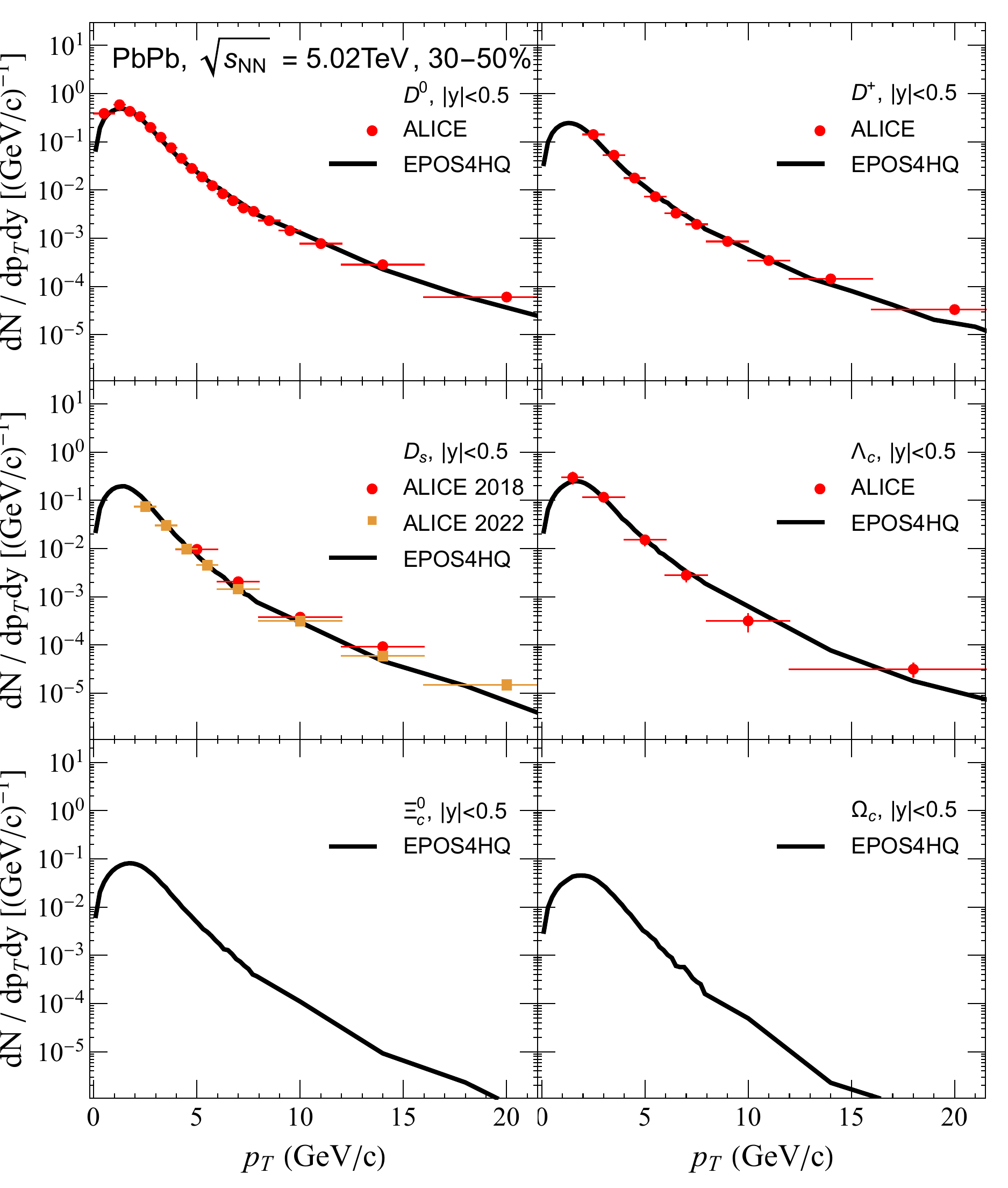}\\
\includegraphics[width=0.46\textwidth]{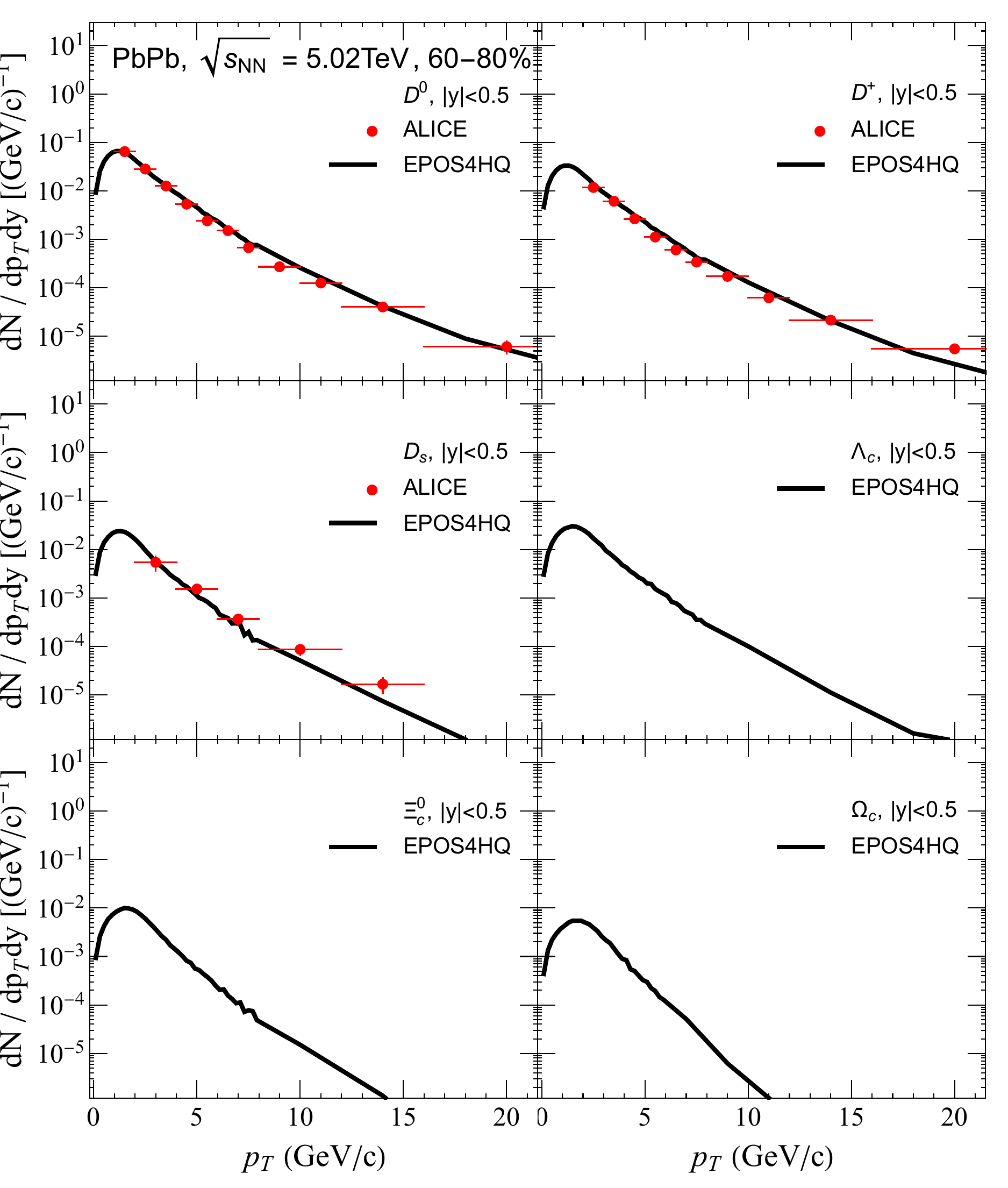}
\caption{$p_T$ spectra of charmed hadrons $D^0$, $D^+$, $D_s$, $\Lambda_c$, $\Xi_c^0$, and $\Omega_c$ in the 30-50\% and 60-80\% centrality classes of PbPb collisions at $\sqrt{s_{\rm NN}}=5.02\rm TeV$. The experimental data of $D$ and $D^+$~\cite{ALICE:2018lyv}, $D_s$~\cite{ALICE:2018lyv}, $\Lambda_c$~\cite{ALICE:2021bib} are from the ALICE.}
\label{fig.spect_PbPb6080}
\end{figure}

The transverse momentum spectra of $D^0$, $D^+$, $D_s$, $\Lambda_c$, $\Xi_c$, and $\Omega_c$ in PbPb collisions at $\sqrt{s_{\rm NN}}=5.02\rm TeV$ and for 0-10\%, 30-50\%, and 60-80\% centrality are shown in Figs~\ref{fig.spect_PbPb010} and~\ref{fig.spect_PbPb6080}.  The EPOS4HQ calculations (thick black line) are compared with the ALICE data: $D^0$ and $D^+$ from ~\cite{ALICE:2021rxa}, $D_s$ form ~\cite{ALICE:2018lyv,ALICE:2021kfc}, and $\Lambda_c$ from ~\cite{ALICE:2021bib}. For $D^0$ ($\Lambda_c$) we show as well the initial $p_T$ distribution of those charm quarks which are finally entrained in $D^0$ ($\Lambda_c$) hadrons (dotted blue line) as well as that before hadronization (dashed green line) to demonstrate how the passage through the QGP and the subsequent hadronization modifies the initial distribution. At high transverse momenta the $p_T$ distribution before the $c$-quarks hadronizes is strongly suppressed as compared to the initial distribution,  testifying the energy loss of the $c$-quark in the QGP.  At low momentum, when the momentum of the $c$-quark is of the order of the averaged momentum of the QGP partons, collisions have the consequence that the $c$-quarks approaches an equilibrium with the QGP. Comparing the dashed green line to the black line, we can see clearly the influence of the hadronization on the transverse momentum change. High $p_T$ charm hadronizes almost exclusively via fragmentation, which shifts the spectra to lower $p_T$. Low $p_T$ charm hadronizes by combining with another light antiquark (or two light quarks for baryons), which shifts the momentum to a higher region. This momentum shift is more pronounced for $\Lambda_c$ due to two light quarks entrained.
Where data are available the transverse momentum distributions of EPOS4HQ are for all centrality bins close to the measured ones.  

\begin{figure}[!htb]
\includegraphics[width=0.46\textwidth]{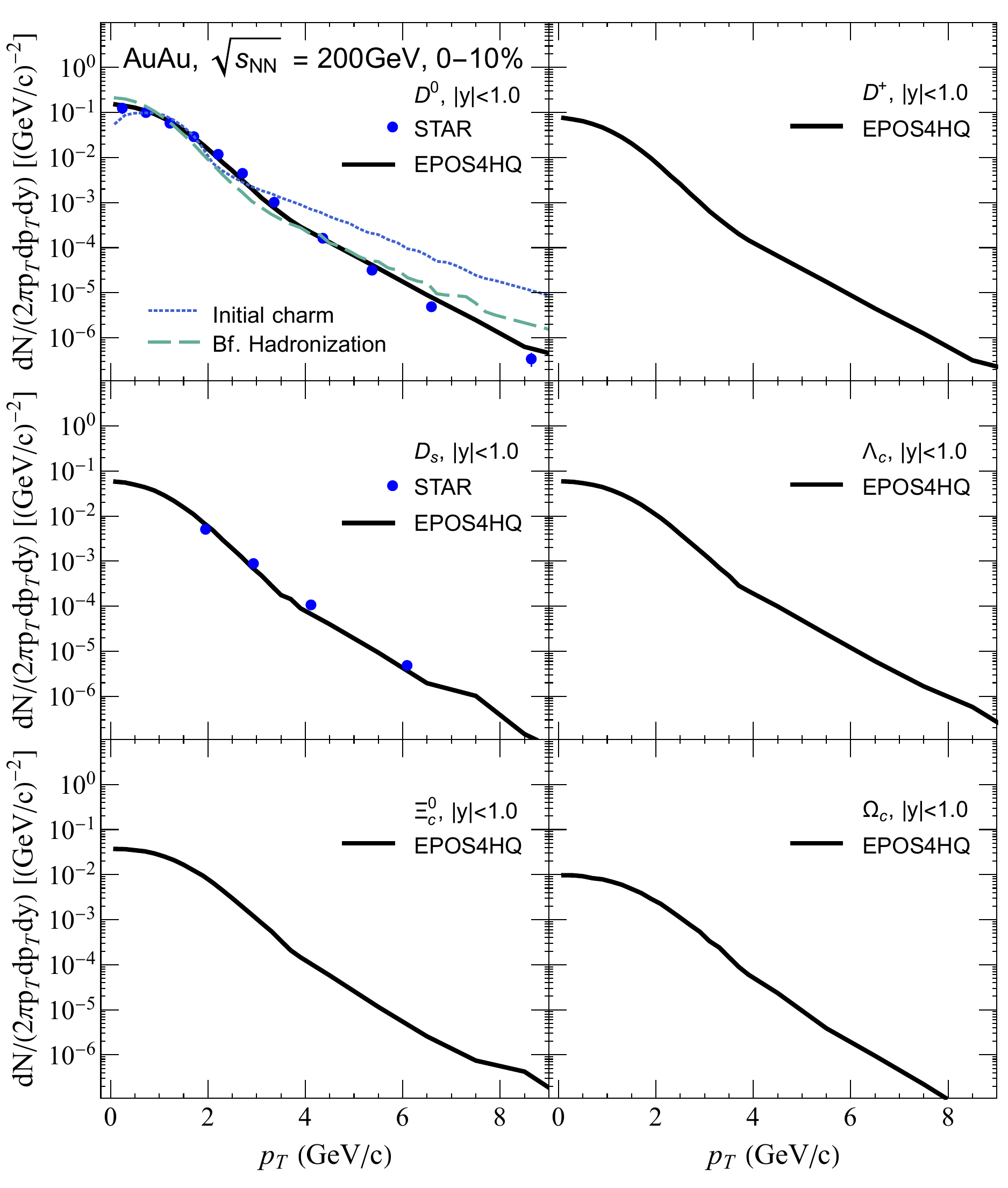}
\includegraphics[width=0.46\textwidth]{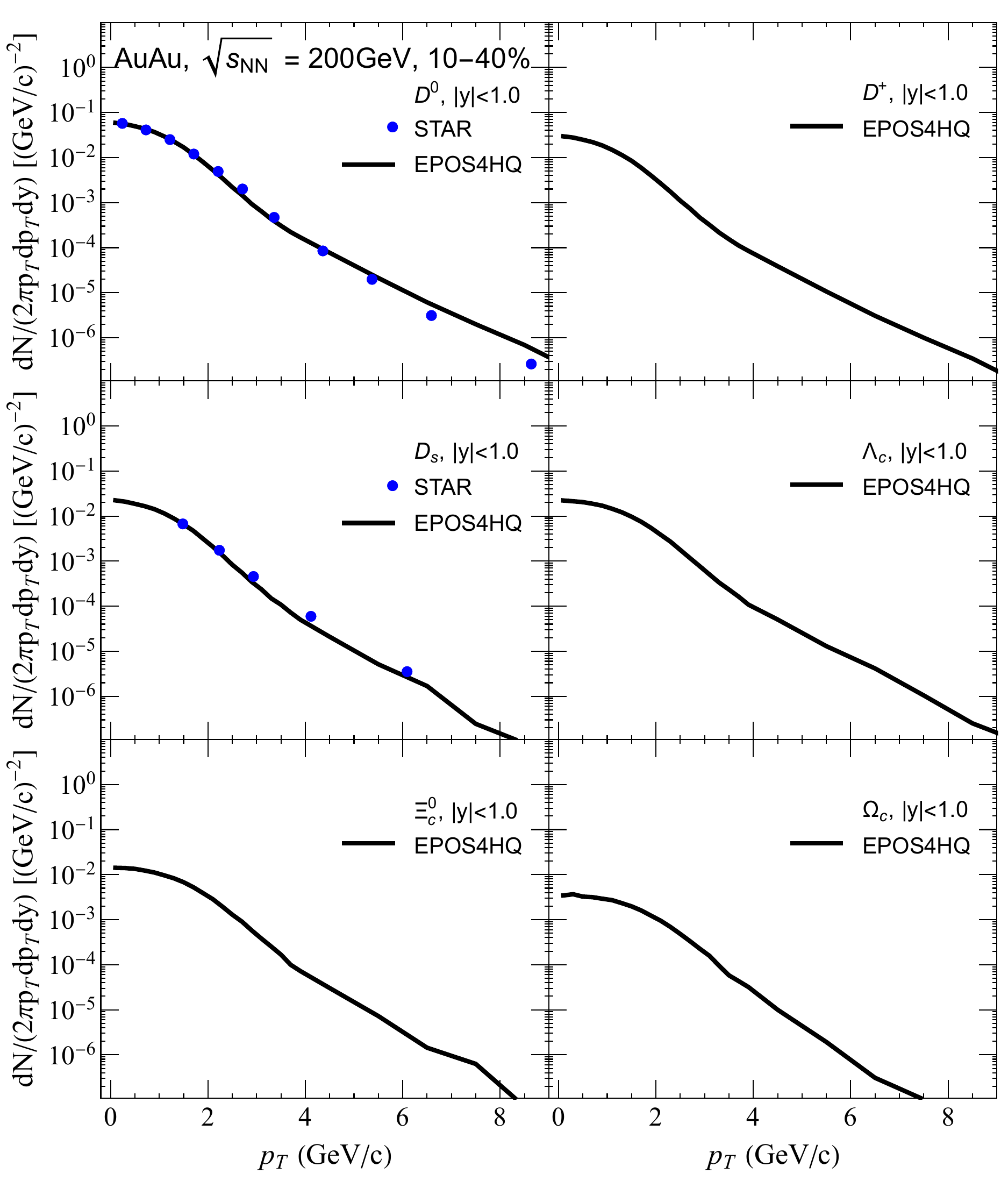}
\caption{$p_T$ spectra of charmed hadrons $D^0$, $D^+$, $D_s$, $\Lambda_c$, $\Xi_c^0$, and $\Omega_c$ in the 0-10\% and 10-40\% centrality classes of AuAu collisions at $\sqrt{s_{\rm NN}}=200\rm GeV$. The experimental data of $D$~\cite{STAR:2018zdy}, $D_s$~\cite{STAR:2021tte} are from the STAR. The thick black line is the EPOS4HQ result, the dotted blue and dashed green lines are the $p_T$ distribution of charm quarks at production and before hadronization, respectively, which finally are entrained in $D^0$ mesons.}
\label{fig.spect_AuAu010}
\end{figure}

The $p_T$ spectra of different charmed hadrons, produced in AuAu collisions at RHIC at $\sqrt{s_{\rm NN}}=200\rm GeV$ and in the 0-10\% and 10-40\% centrality bins are shown in Figs.~\ref{fig.spect_AuAu010}.  The EPOS4HQ results (thick black line) are compared with the STAR data 
\cite{STAR:2018zdy} for $D$-mesons  and ~\cite{STAR:2021tte} for $D_s$  mesons. For the $D^0$ data we show as well the initial $p_T$ distribution of those charm quarks which are finally entrained in a $D^0$ meson (dotted blue line) as well as that before hadronization (dashed green line) to demonstrate how the passage through the QGP and the subsequent hadronization modifies the initial distribution. As compared to the LHC 
results the energy loss while traversing the QGP shifts the momentum less and therefore the spectrum at high $p_T$ is less suppressed.  Also at RHIC we reproduce nicely the available experimental data.

\subsubsection{bottom hadrons}
\begin{figure}[!htb]
\includegraphics[width=0.46\textwidth]{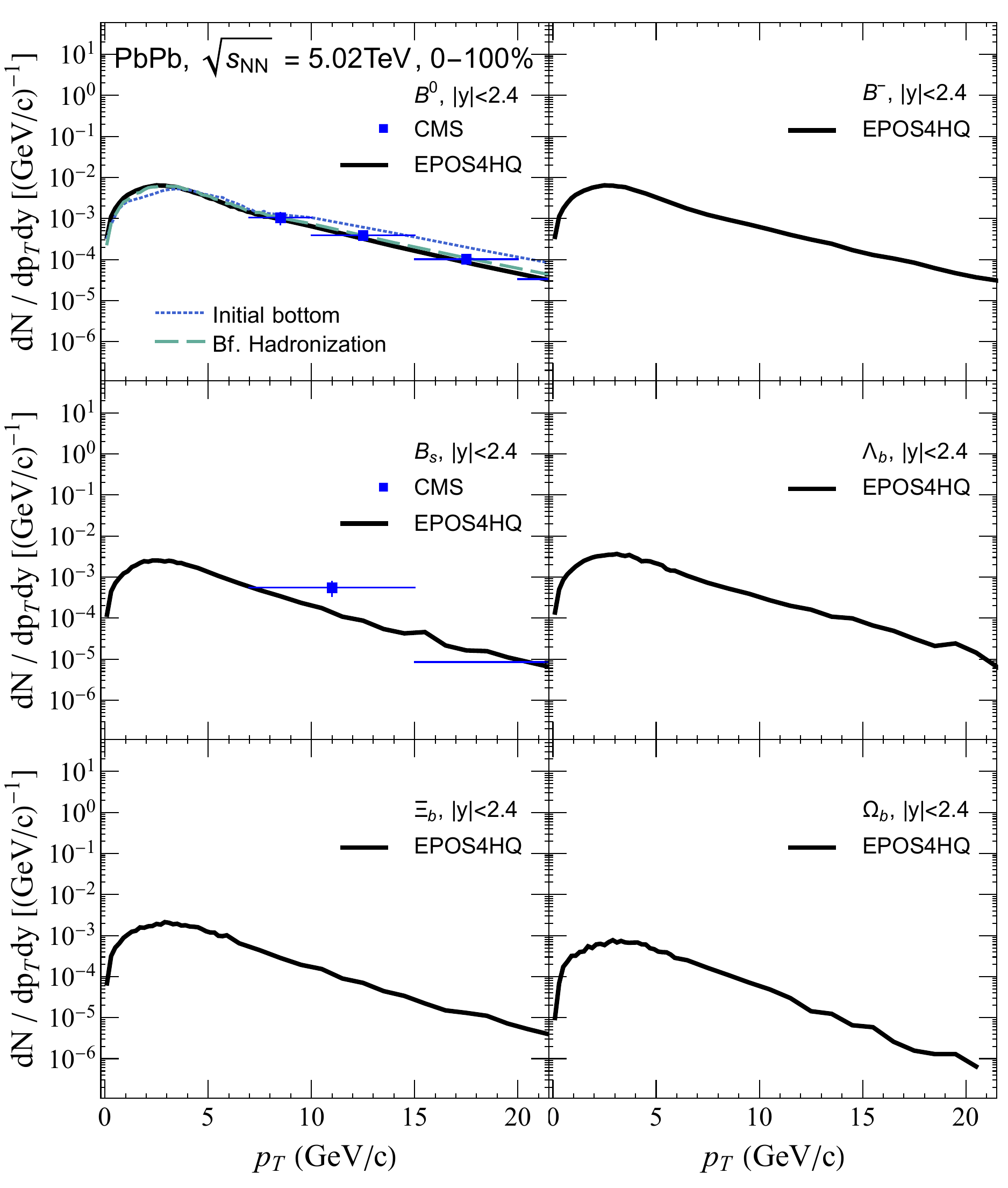}
\caption{$p_T$ spectra of charmed hadrons $B^0$, $B^-$, $B_s$, $\Lambda_b$, $\Xi_b$, and $\Omega_b$ in the 0-100\% centrality class of PbPb collisions at $\sqrt{s_{\rm NN}}=5.02\rm TeV$. The experimental data of $B^0$~\cite{CMS:2017uoy} and $B_s$~\cite{CMS:2018eso} are from the CMS collaboration. The thick black line is the EPOS4HQ result, the dotted blue and dashed green lines are the $p_T$ distribution of bottom quarks at production and before hadronization, respectively, which finally are entrained in $B^0$ mesons.}
\label{fig.spect_PbPb0100_b}
\end{figure}

The transverse momentum spectra of bottom hadrons $B^0$, $B^-$, $B_s$, $\Lambda_b$, $\Xi_b$, and $\Omega_b$ in the 0-100\% centrality class of PbPb collisions at $\sqrt{s_{\rm NN}}=5.02\rm TeV$ are shown in Fig. \ref{fig.spect_PbPb0100_b}. The experimental data of $B^0$~\cite{CMS:2017uoy} and $B_s$~\cite{CMS:2018eso} are from the CMS collaboration. For the $B^0$ data we show as well the initial $p_T$ distribution of the charm quarks (dotted blue line) as well as that before hadronization (dashed green line), which are finally part of $B^0$ mesons. The momentum loss of a bottom quark while passing through the plasma is less than that of a charm quark. This is due to the kinematics of the interaction of the large mass $b$-quarks with the QGP partons, which suppresses the momentum transfer as compared to the interaction of $c$-quarks. 

\begin{figure}[!htb]
\includegraphics[width=0.4\textwidth]{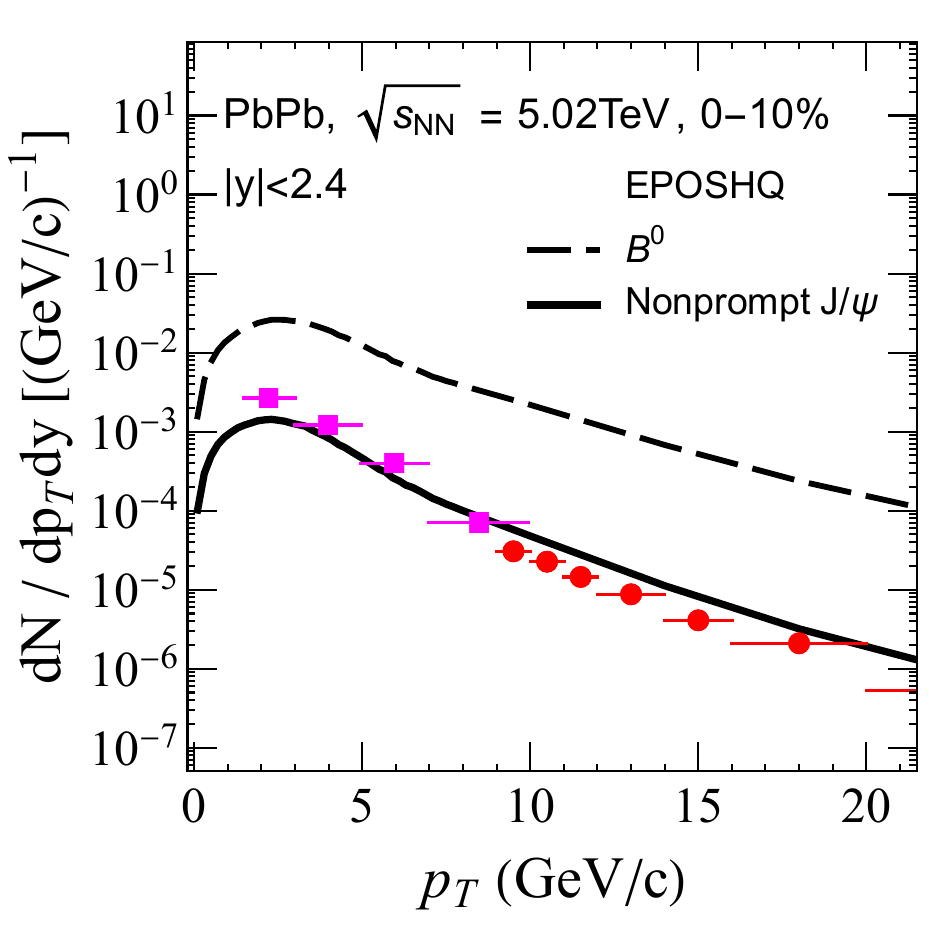}
\caption{$p_T$ spectrum of $B^0$ and nonprompt $J/\psi$ in the 0-10\% PbPb collisions at 5.02TeV. The experimental data are from the ATLAS~\cite{ATLAS:2018hqe} and ALICE~\cite{ALICE:2023hou}.} 
\label{fig.spect_nonpromptjpsi}
\end{figure}

Another source of information about $B$-mesons are the decay products, especially the non-prompt $J/\psi$s. Their $p_T$ spectrum has been measured for central events in PbPb at $\sqrt{s_{\rm NN}}$  = 5.02 GeV by the ATLAS \cite{ATLAS:2018hqe} and the ALICE~\cite{ALICE:2023hou} collaborations. Their results are displayed in Fig.  \ref{fig.spect_nonpromptjpsi} and compared to that of EPOS4HQ calculations.  We display in this figure as well the $p_T$ distribution of the $B^0$ mesons in this centrality bin.  In our calculation we assumed that non-prompt $J/\psi$ come exclusively from $B^0$ decays and that the experimental $p_T$ differential branching ratio $\mathcal{B}(p_T)$ is the same in pp as in PbPb. Then we can calculate the non-prompt $J/\psi$ $p_T$ spectrum in PbPb by multiplying the calculated $B^0$($p_T$) spectrum in PbPb with $\mathcal{B}(p_T)$. Doing this we assume that shadowing does not influence the branching ratio, an assumption which creates an additional uncertainty.


\subsection{Yield ratio}
\begin{figure}[!htb]
\includegraphics[width=0.46\textwidth]{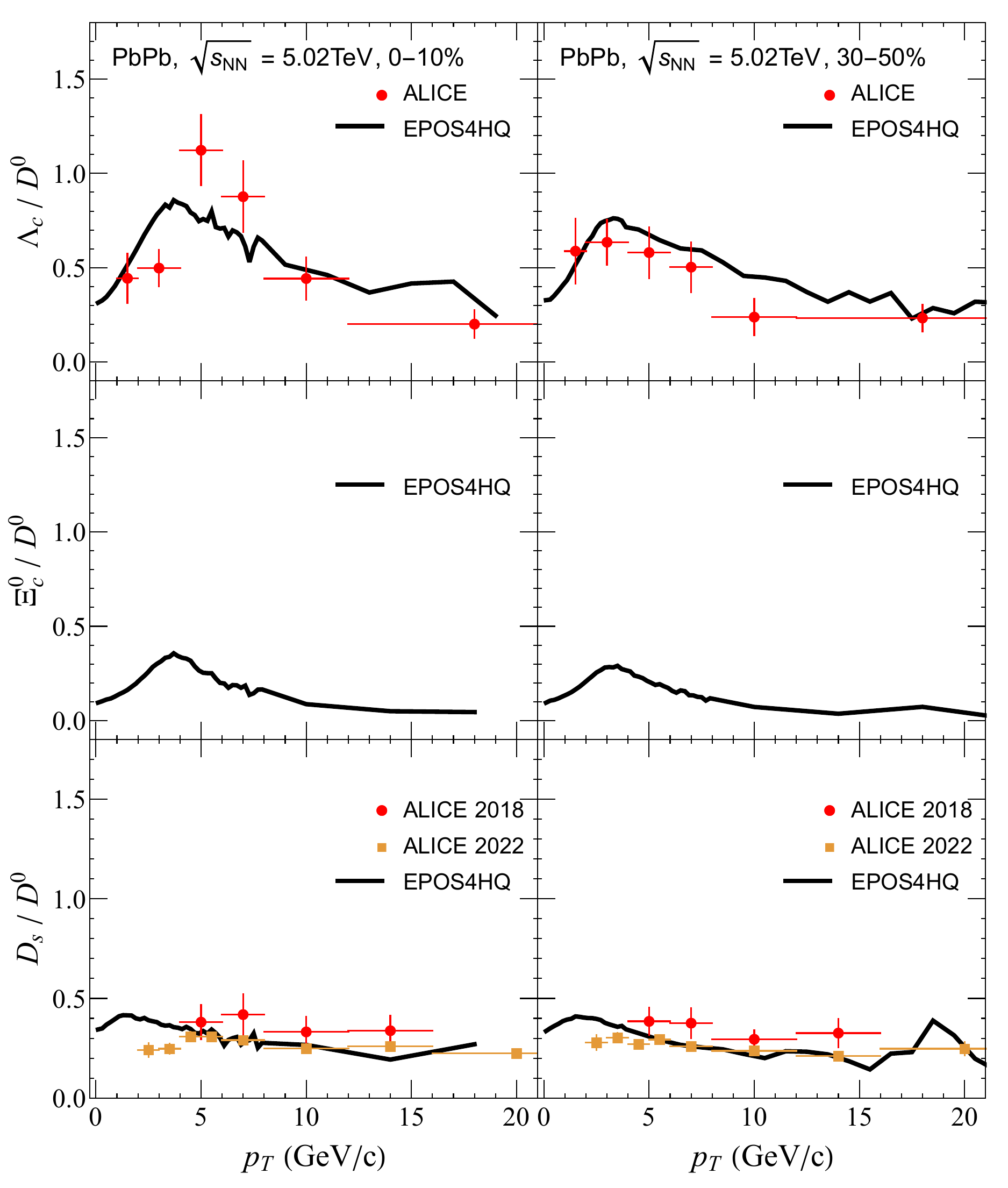}
\caption{Yield ratio of $\Lambda_c/D^0$, $\Xi_c^0/D^0$, and $D_s/D^0$ in the 0-10\% (left) and 30-50\% (right) centrality class of PbPb collisions at $\sqrt{s_{\rm NN}}=5.02 \rm TeV$. The experimental data are from the ALICE, $\Lambda_c/D^0$~\cite{ALICE:2021bib}, $D_s/D^0$~\cite{ALICE:2018lyv,ALICE:2021kfc}.}
\label{fig.ratio_PbPb010}
\end{figure}
Next we present the yield ratio between different heavy flavor hadrons as a function of $p_T$. 
Yield ratios of charmed hadrons are important because they present an experimental probe to study
fragmentation functions into the different hadron species. This is especially true at high $p_T$ where on the one side hadrons are almost exclusively produced by fragmentation and on the other side hardly any theoretical calculation of these fragmentation functions is available.  Yield ratios can also eliminate systematic uncertainties and have therefore a smaller error than absolute yields. They may also contribute to solve physics questions.  For example, the yield ratio of $D_s/D^0$ can reveal the strangeness enhancement in the QGP phase of the  heavy-ion collisions. The enhancement of the baryon to meson ratio, as compared to the $e^+e^-$  collisions, e.g. $p/\pi$, $\Lambda/K$, shows the importance of hadronization by recombination which is absent in $e^+e^-$ collision where hadronization is exclusively described by fragmentation functions.  This enhancement has also been observed in the heavy flavour sector, like in the $\Lambda_c/D^0$ ratio, and is even present in high energy pp collisions  \cite{STAR:2019ank,ALICE:2021bib}.

In  Fig~\ref{fig.ratio_PbPb010}  we present the yield ratios of $\Lambda_c/D^0$, $\Xi_c^0/D^0$, and $D_s/D^0$ in the 0-10\%  (left) and 30-50\% (right) centrality classes of PbPb collisions at $\sqrt{s_{\rm NN}}=5.02 \rm TeV$. The experimental data are from the ALICE collaboration, $\Lambda_c/D^0$ from ~\cite{ALICE:2021bib}, $D_s/D^0$ from ~\cite{ALICE:2018lyv,ALICE:2021kfc}. We observe that in both centrality classes the experimental $D_s/D^0$ ratio is well reproduced by the EPOS4HQ calculations. This is a strong hint that in the QGP, from where the light quarks originate, the strange quark multiplicity corresponds to the equilibrium value for a system close to $T_{\rm FO}$, as assumed in the EPOS4HQ approach. Also the $\Lambda_c/D^0$ ratio of EPOS4HQ is
for both centrality classes and in the whole $p_T$ interval in agreement with experiment~\cite{ALICE:2021bib} , including the peak structure of the distribution at around $p_T = 3 $ GeV. This ratio is well above the ratio measured in $e^+e^-$ collisions. The enhancement of the baryon to meson ratio in heavy ion collisions can be well understood within the quark recombination/coalescence model~\cite{Oh:2009zj,Plumari:2017ntm,Zhao:2018jlw,Cao:2019iqs,He:2019vgs} and therefore the agreement between theory and data gives evidence that at low $p_T$ the majority of $\Lambda_c$ is created in coalescence processes. At higher $p_T$, where fragmentation dominates, the agreement shows that the employed fragmentation functions \cite{Braaten:1994bz,Cacciari:2005rk} are realistic. We also display the prediction for the $\Xi_c/D^0$ ratio. The shape is similar to the $\Lambda_c/D^0$, but the value is reduced to one half of the former due to the strange quark.

\subsection{Nuclear modification factor $R_{AA}$}

\begin{figure}[!htb]
\includegraphics[width=0.4\textwidth]{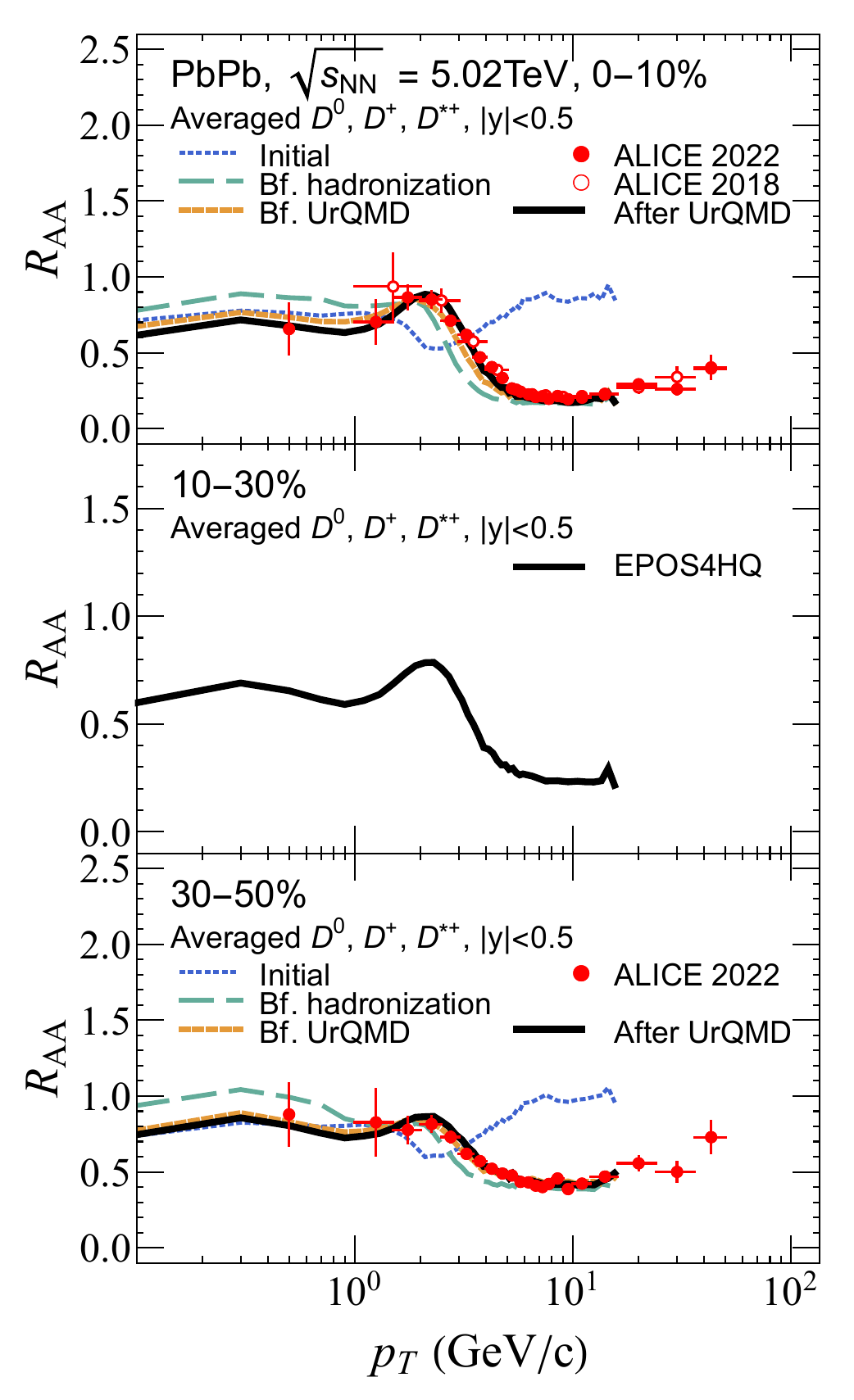}
\caption{Nuclear modification factor $R_{AA}$ of $D^0$ in the 0-10\%, 10-30\%, and 30-50\% PbPb collisions at 5.02TeV. The experimental data are from the ALICE~\cite{ALICE:2018lyv,ALICE:2021rxa}. The black thick (orange lines) lines
are the ratio of $D^0$ in PbPb and pp after (before) UrQMD. The blue dotted and green dashed lines are the ratio of charm quarks at production
and before hadronization, respectively, which finally are entrained in $D^0$ mesons.} 
\label{fig.Raa_PbPb}
\end{figure}

\begin{figure}[!htb]
\includegraphics[width=0.4\textwidth]{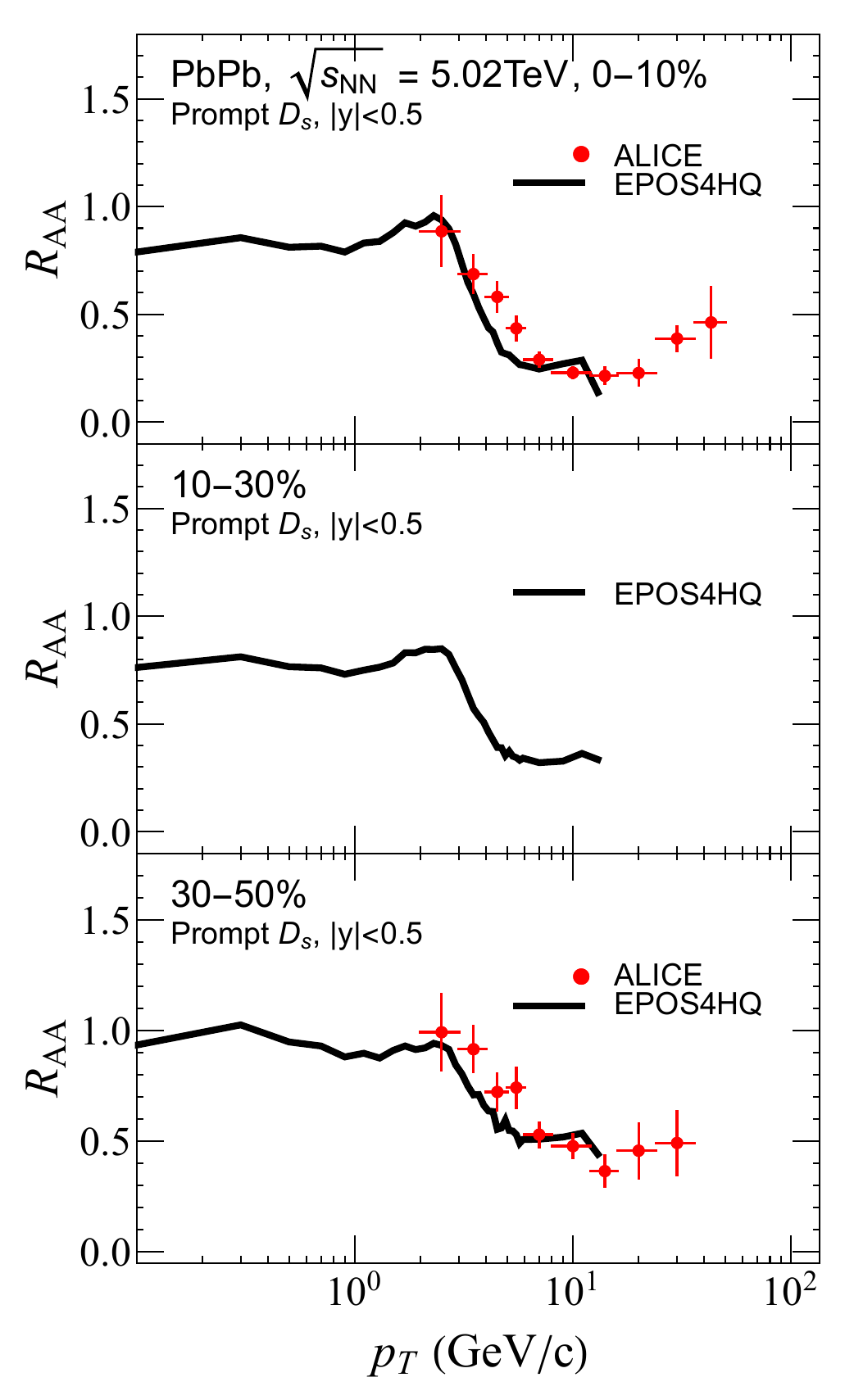}
\caption{Nuclear modification factor $R_{AA}$ of $D_s$ in the 0-10\%, 10-30\%, and 30-50\% PbPb collisions at 5.02TeV. The experimental data are from the ALICE~\cite{ALICE:2021kfc}.} 
\label{fig.Raa_PbPb_Ds}
\end{figure}
\begin{figure}[!htb]
\includegraphics[width=0.4\textwidth]{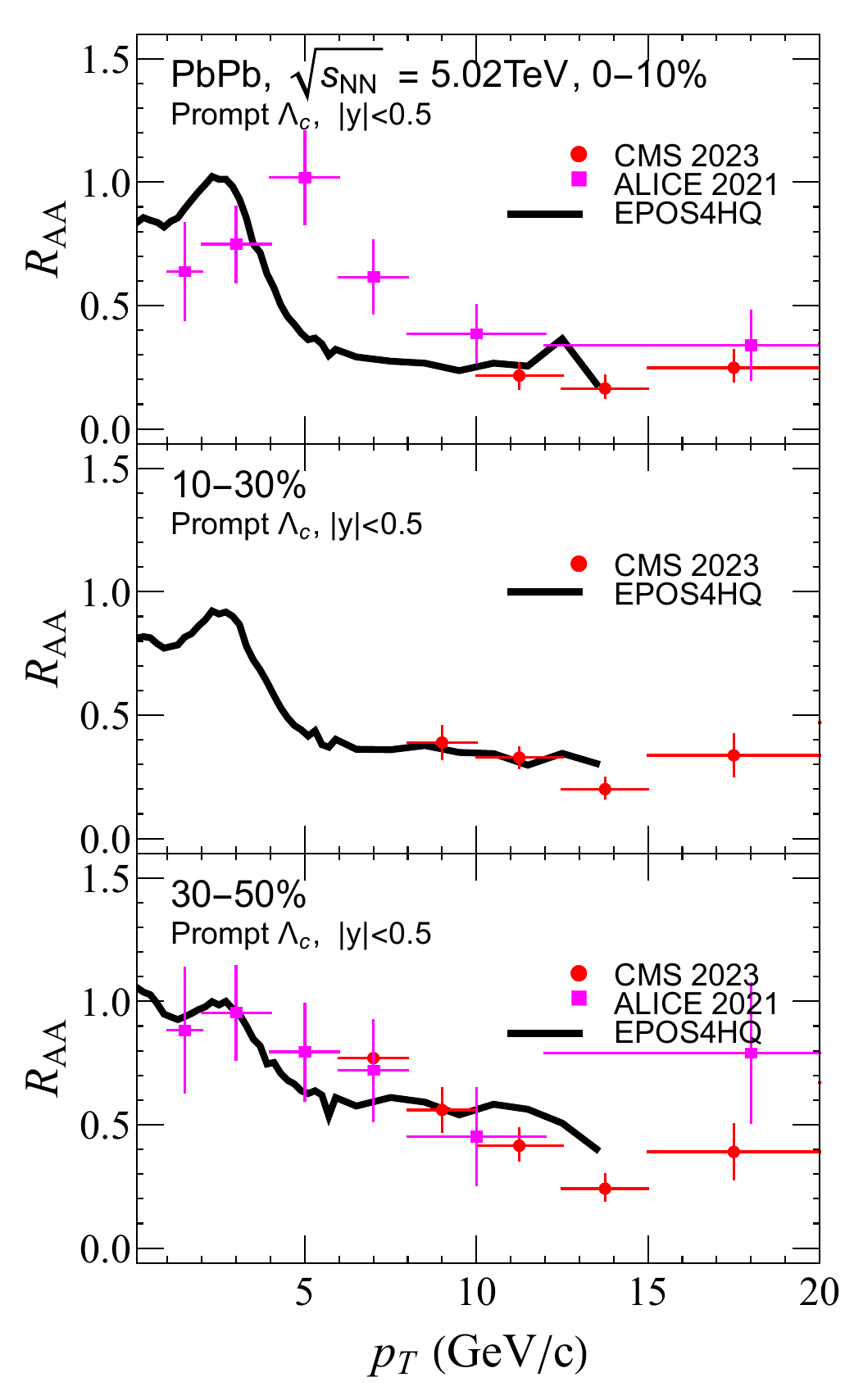}
\caption{Nuclear modification factor $R_{AA}$ of $\Lambda_c$ in the 0-10\%, 10-30\%, and 30-50\% PbPb collisions at 5.02TeV. The experimental data are from the ALICE~\cite{ALICE:2021bib} and CMS~\cite{CMS:2023frs}.} 
\label{fig.Raa_PbPb_Lc}
\end{figure}

Heavy quarks are produced in primary hard scattering processes, which occur in the early stage of the heavy-ion collisions.  Therefore it is useful  to define the nuclear modification factor 
\begin{eqnarray}
R_{AA}\equiv{1\over  \langle N_{\rm coll} \rangle} {dN_{AA}/dp_T \over dN_{pp}/dp_T}.
\end{eqnarray} 
which compares the $p_T$ distribution in AA collisions with that in pp collisions  multiplied by 
$\langle N_{\rm coll} \rangle$, the average number of initial hard binary collisions. If the $p_T$ distribution, observed in heavy-ion collisions, is just a superposition of that observed in
pp collisions we expect $R_{AA}=1$. In Fig.~\ref {fig.Raa_PbPb} the full black lines show the $R_{AA}$ for the finally observed hadrons, calculated in the EPOS4HQ approach, and  
the experimental data ~\cite{ALICE:2018lyv,ALICE:2021rxa} are marked as red points. We see that $R_{AA}$ is smaller than 1.  This has two reasons: 
\begin{itemize}
\item Nuclear shadowing modifies the parton distribution function of a nucleus in comparison to that of a proton. In \cite{ALICE:2021rxa} the measured  integrated $R_{AA}$ for $D^0$ mesons (0.689 for the centrality [0-10\%] and for $|y|<0.5$, what EPOS4HQ reproduces within the error bars) is compared with perturbative QCD calculations of $D^0$-meson production including only initial-state effects modeled using two different sets of nuclear parton distribution functions (PDF), namely nCTEQ15 and EPPS16. These calculations show, however with a large error bar, that the major fraction of the suppression comes from the PDF's, i.e. from the fact that in a nucleus the parton distribution function is different than in a proton. This is confirmed by EPOS4HQ. The $R_{AA}$ for the initially produced charm quarks as around 1 at large $p_T$ but is almost constant $\sim0.7$ at low $p_T$ values, as seen in Fig.~\ref {fig.Raa_PbPb}, top, as a blue dotted line. There we display $R_{AA}$ as a function of $p_T$ for [0-10\%] central PbPb collisions at $\sqrt{s_{\rm NN}}$ = 5.02 TeV. After production, heavy quarks propagate in the QGP and interact with the thermal partons by exchanging energy and momentum via elastic and inelastic collisions, as discussed in section III. Due to the energy loss, at high $p_T$ the initial momentum of a $c$-quark will shift to a lower momentum, the spectrum becomes softer and $R_{AA}$ falls well below one (green dashed line).  At low $p_T$ the momentum of the $c$-quark increases due to these collisions and the $c$-quarks approaches thermal equilibrium. This leads to a slight increase of $R_{AA}$. 
\item Second, due to hadronization by coalescence, which is more important in AA than in pp collisions, we observe at low $p_T$ an increased production of heavy baryons and therefore a lower production of heavy mesons. This lowers $R_{AA}$ at low $p_T$, as seen, when comparing the green dashed and the orange line. 
\end{itemize}
Both processes lead to a complex structure of $R_{AA}$ with a maximum at intermediate $p_T$. For less central collisions, shown in Fig.~\ref {fig.Raa_PbPb} in the middle for [10-30\%] and in the bottom row for [30-50\%] centrality, the form of $R_{AA}$ as a function of $p_T$ is similar but $R_{AA}$ increases because the energy loss in the plasma gets smaller. The $R_{AA}$ of $D_s$ and $\Lambda_c$ are also investigated and shown in Figs.~\ref{fig.Raa_PbPb_Ds} and~\ref{fig.Raa_PbPb_Lc}, which show a good agreement with the experimental data.

\begin{figure}[!htb]
\includegraphics[width=0.4\textwidth]{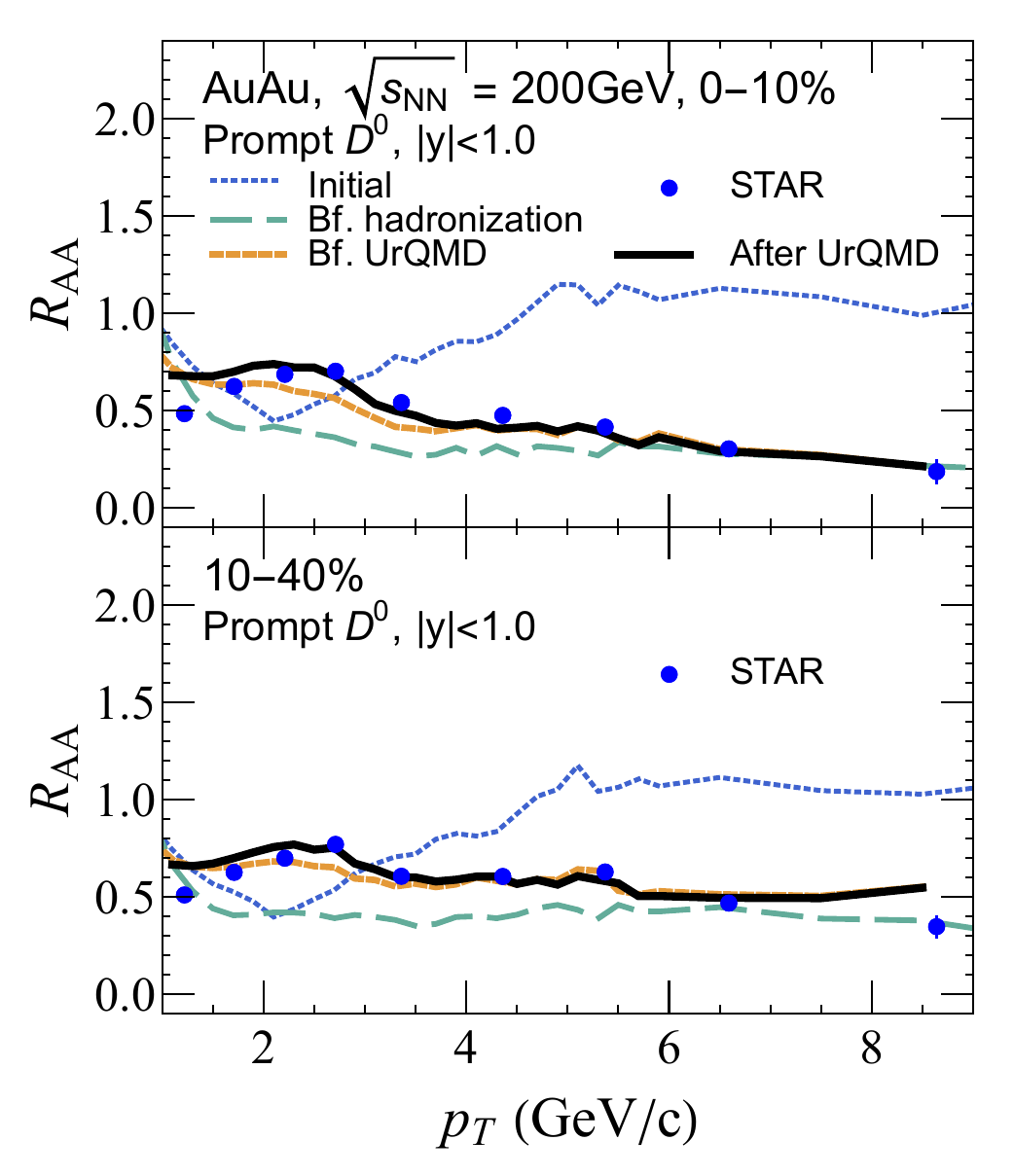}
\caption{Nuclear modification factor $R_{AA}$ of $D^0$ in the 0-10\% and 10-40\% AuAu collisions at 200GeV. The experimental data are from the STAR~\cite{STAR:2018zdy}. The black thick (orange lines) lines
are the ratio of $D^0$ in AuAu and pp after (before) UrQMD. The blue dotted and green dashed lines are the ratio of charm quarks at production
and before hadronization, respectively, which finally are entrained in $D^0$ mesons.} 
\label{fig.Raa_AuAu}
\end{figure}

Fig.~\ref{fig.Raa_AuAu} shows the centrality dependence of $R_{AA}$ for $D^0$-mesons in AuAu collisions at $\sqrt{s_{\rm NN}}= $ 200 GeV.  We display
$R_{AA}(p_T)$ for two different centrality classes [0-10\%] and [10-40\%] and for the $p_T$ interval for which experimental data are available and not only extrapolations. The STAR data are from~\cite{STAR:2018zdy}. In this figure we display as well  the $R_{AA}$ of the those $c$-quarks quarks which are later part of $D^0$ mesons.  The blue dotted line shows $R_{AA}$  at the moment when the $c$-quarks are produced.  Also at RHIC energies for large $p_T$ $c$-quarks  $R_{AA}$ is equal one because shadowing modifies only the $c$-quark distribution at low $p_T$. There the suppression of $c$-quarks in AuAu can reach 50\% as compared to pp. The interaction of heavy quarks with the QGP (difference between green long dashed line and the blue dotted line)  modifies strongly the $R_{AA}$ at high $p_T$. Coalescence dominates for low $p_T$ heavy quarks and therefore the momentum of the $D$-meson is larger than that of the heavy $c$-quark embedded, leading to an enhancement of $R_{AA}$  at small $p_T$ (difference between dashed orange and long dashed green line) . The final hadronic interactions (difference between black and long dashed green line) have little influence of the final form of $R_{AA}(p_T)$.

\begin{figure}[!htb]
\includegraphics[width=0.4\textwidth]{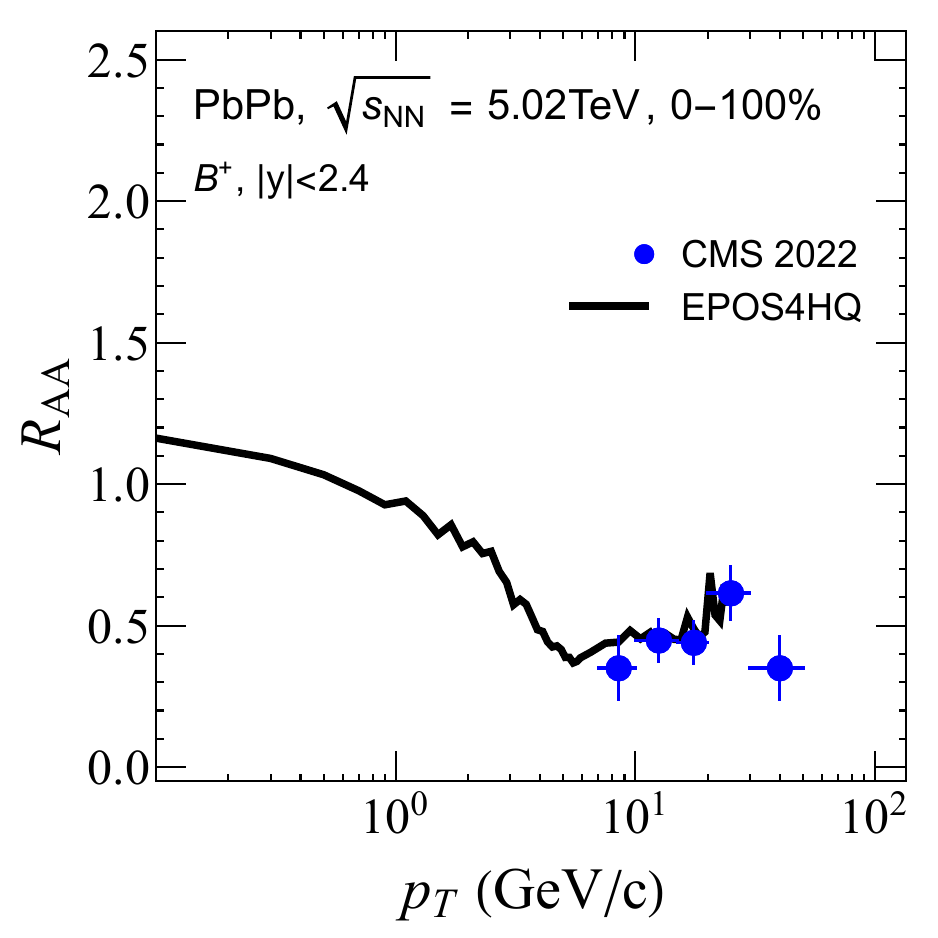}
\caption{Nuclear modification factor $R_{AA}$ of $B^+$ in the 0-100\% PbPb collisions at 5.02TeV. The experimental data is from the CMS collaboration~\cite{CMS:2017uoy}.} 
\label{fig.Raa_B}
\end{figure}

Finally we compare in Fig.~\ref{fig.Raa_B} $R_{AA}(p_T)$ of $B^+$ mesons, measured by the CMS collaboration for PbPb collisions at $\sqrt{s_{\rm NN}}$ = 5.02 TeV with the EPOS4HQ calculations. $B$-mesons are too heavy to approach equilibrium with the QGP. For this reason we observe
only a momentum shift of the $b$-quarks towards lower $p_T$ values.
The data agree quite nicely with the EPOS4HQ calculations.  

From the confrontation of the EPOS4HQ results with calculations employing the generalized parton distribution function we
can conclude that the shadowing in EPOS4HQ is well described. From the agreement of the EPOS4HQ results with data we can furthermore conclude that also the other processes, which influence the final $p_T$ distributions of open heavy flavour mesons,
are quite reasonable described.

\subsection{Elliptic flow $v_2$}

The elliptic flow $v_2$  is another key observable to study the interaction of heavy quarks with the QGP. 
In non-central heavy ion collisions, the initial spatial anisotropy of the overlap region is converted into an anisotropic azimuthal distribution in momentum space of the final particles at low $p_T$. This anisotropy can be characterized in terms of Fourier coefficients 
\begin{equation}
v_n \equiv \langle \cos[n(\Phi-\Psi_n)]\rangle,    
\label{eq:v2}
\end{equation} 
where $\Phi$ is the azimuthal angle of the particle and $\Psi_n$ is the azimuthal angle of the event plane for the nth-order harmonics. Elliptic flow, $v_2$, is the second-order coefficient, whose value is in hydrodynamical calculations proportional to the initial spatial eccentricity of the overlap region. 
Being produced in a hard process, at creation heavy quarks have $v_2=0$. So any observed $v_2$ of heavy mesons is due to the interaction of heavy quarks with QGP partons or of heavy mesons with other hadrons. The final elliptic flow of heavy flavor hadrons  comes from the heavy quark itself and also from  the light antiquark, which merges with the heavy quark in the hadronization process. Also the hadronization process itself
may contribute slightly. The measurement of $v_2$ of heavy-flavor hadrons at low $p_T$ can therefore help to quantify the coupling between heavy quarks and the hot medium as well as to understand the hadronization mechanism. At high $p_T$, heavy quarks hadronize via fragmentation. In this case, the elliptic flow $v_2$  together with $R_{AA}$ can be used to explore the path-length dependence of the  in-medium energy loss as in the almond shaped overlap region the path length of the heavy quark in the QGP medium depends on its azimuthal direction. 
\begin{figure}[!htb]
\includegraphics[width=0.4\textwidth]{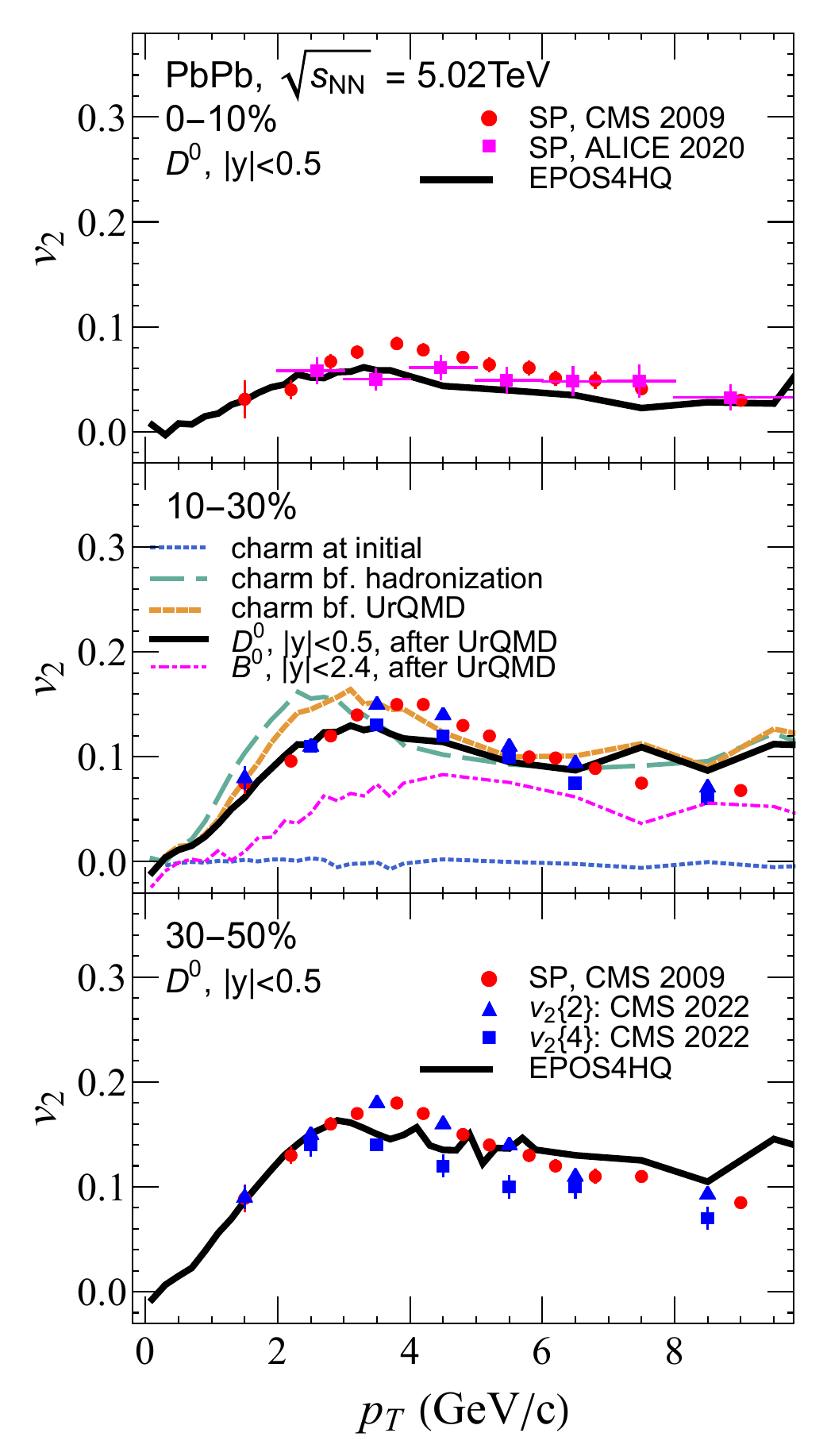}
\caption{Elliptic flow $v_2$ of $D^0$ in the 0-10\%, 10-30\%, and 30-50\% centrality classes of PbPb collisions at $\sqrt{s_{\rm NN}}=5.02 \rm TeV$. The experimental data are from the ALICE~\cite{ALICE:2020iug} and CMS~\cite{CMS:2020bnz,CMS:2021qqk} collaborations. The black thick (orange lines) lines
are the $v_2$ of $D^0$ after (before) UrQMD. The blue dotted and green dashed lines are $v_2$ of charm quarks at production
and before hadronization, respectively. The magenta dashed-dotted line in the middle panel is the $v_2$ of $B^0$ after UrQMD.}
\label{fig.v2_PbPb}
\end{figure}

\begin{figure}[!htb]
\includegraphics[width=0.4\textwidth]{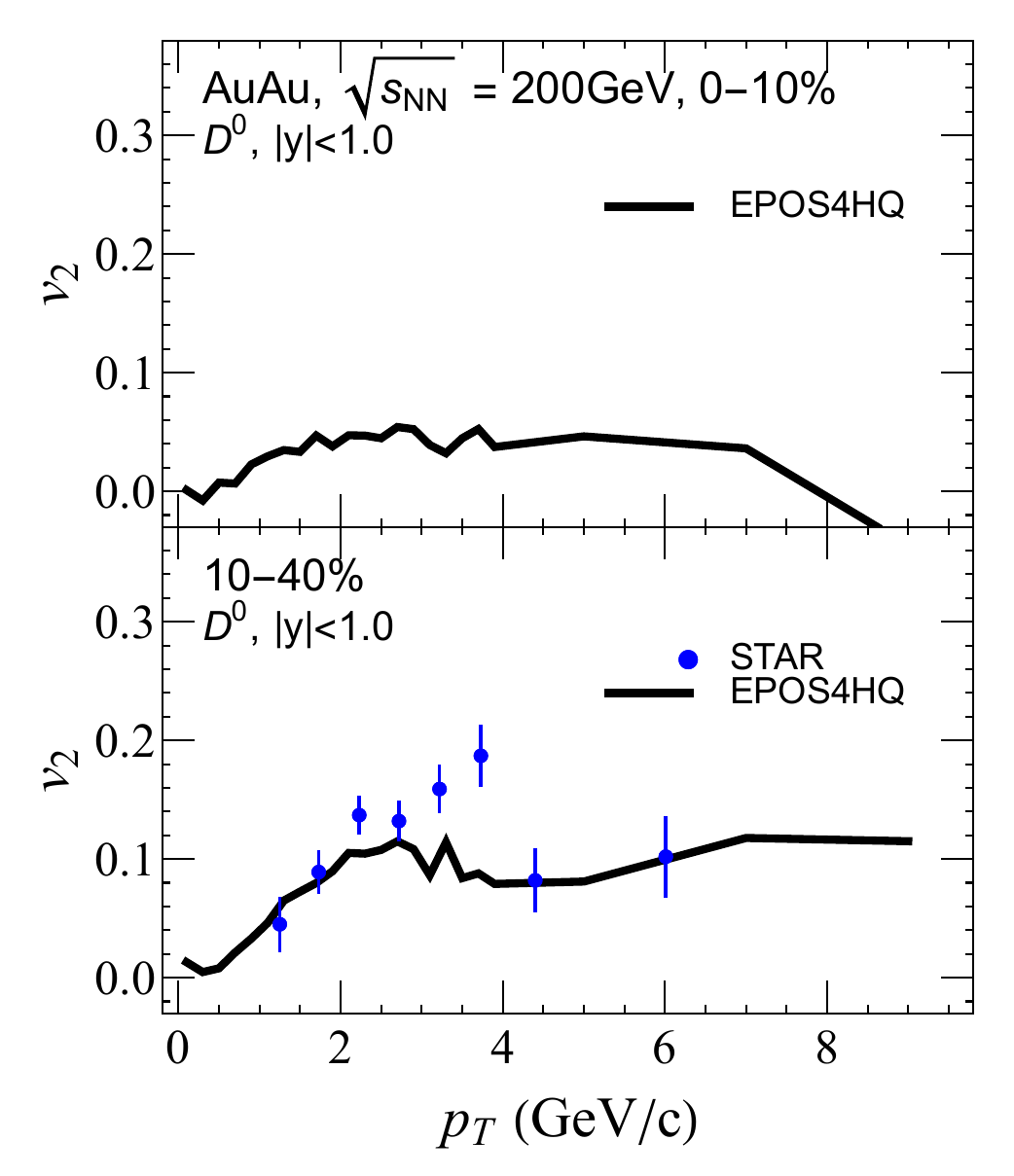}
\caption{Elliptic flow $v_2$ of $D^0$ in the 0-10\% and 10-40\% centrality classes of AuAu collisions at $\sqrt{s_{\rm NN}}=200 \rm GeV$. The experimental data are from the STAR~\cite{STAR:2017kkh}.}
\label{fig.v2_AuAu}
\end{figure}

In EPOS4 the event plane $\Psi_n$ can be determined, what is not possible in experiments with a limited acceptance. Therefore a multi particle correlation or cumulants method has been developed to determine $v_2$. Depending on the number n of particles, whose correlations are calculated, the result is named $v_2\{n\}$. The higher the order, the more
non-flow effects, such as resonance decays and jets, are eliminated.
This technique has been widely applied in the light-flavor sector~\cite{CMS:2012zex}. Recently it has also been used for the heavy flavour flow analysis ~\cite{CMS:2021qqk}. 

In Fig.~\ref{fig.v2_PbPb} we show the $p_T$ dependence of  $v_2\{2\}$ and $v_2\{4\}$ of $D^0$ mesons measured by the CMS collaboration~\cite{CMS:2021qqk} at midrapidity for PbPb collisions at $\sqrt{s_{\rm NN}}$ = 5.02 TeV in different centrality bins. From top to bottom we display 
the results for the centrality bins [0-10\%], [10-30\%] and [30-50\%]. We see  that indeed the experimental $v_2\{4\}$ is slightly smaller than  $v_2\{2\}$. In older publications usually $v_2\{2\}$ or the equivalent scalar product (SP) approach has been used. The EPOS4HQ results, obtained with Eq.~\eqref{eq:v2} and shown as well, should be compared with $v_2\{4\}$.

We display in this figure as well the EPOS4HQ results for different times during the evolution of the heavy-ion reaction. Initially the flow of heavy quarks (dotted blue line) is compatible with zero. After passing through the QGP (dashed green line) the $c$-quarks have acquired already $v_2$ by collisions with the QGP partons. Hadronization does not change $v_2$ substantially as can be concluded by comparing the orange short-dashed line ($v_2$ of $D^0$ mesons after hadronization) with the green dashed line but shifts the maximum to larger $p_T$ values, mainly because the $p_T$ of the $D$-meson is different from that of the c-quark. Hadronic interactions (the difference between the orange and the black line) still change slightly the value of $v_2(p_T)$. The final distribution is given by the black line which agrees quite nicely with the experimental data.
For the [30-50\%] centrality bin we show in Fig.~\ref{fig.v2_PbPb} as well $v_2$ of the $B^0$ mesons, as a magenta dashed-dotted line, for which no data exists. At low $p_T$ it is considerable smaller than that of the $D^0$ mesons because due to their large mass the transfer of $v_2$ to $b$-quarks during the interactions with the QGP partons is less efficient. At higher $p_T$, where the $v_2$ is created by the path length difference, $v_2$ of $B^0$ mesons approaches that of $D^0$ mesons because the mass difference gets increasingly less important.  

At RHIC $v_2(p_T)$ of $D^0$ mesons has been measured by the STAR collaboration ~\cite{STAR:2017kkh}. The data for semicentral AuAu collisions at $\sqrt{s_{\rm NN}}=$ 200 GeV are displayed in Fig.~\ref{fig.v2_AuAu} and compared with EPOS4HQ results. We display in this figure also our results for central collisions for which no data are available yet. 

\section{Summary}
\label{sec.summary}
We used the newly developed EPOS4 approach, which describe quite well a multitude of light hadron observables, to study the physics of heavy quarks. For this we implemented the same
description of the dynamics of heavy quarks in the QGP as in the former EPOS3/EPOS2 calculations and included a new hadronization approach, which allows to create in the hadronization process all known heavy hadron species. This approach is named EPOS4HQ. In this approach we investigated the production of heavy hadrons in ultrarelativistic heavy-ion collisions at RHIC and LHC energies. We compared transverse momentum spectra, yield ratio, nuclear modification and elliptic flow of different charmed hadrons for both PbPb collisions at $\sqrt{s_{\rm NN}}$ = 5.02 TeV and AuAu collisions at $\sqrt{s_{\rm NN}}$ = 200 GeV. As already in pp collisions, we find a
quite good agreement of all observables with the available experimental data. This agreement is considerable better than for the results we obtained using the same description of the heavy quark/heavy meson dynamics in EPOS2 and EPOS3. We can also extend this approach to RHIC energies what was not possible
with the EPOS3/EPOS2 approach.

We studied as well in detail how the different observables
are modified during the heavy ion collisions. We displayed the
initial distribution of $c$-quarks, the modification of the $c$-quark distribution due to heavy quark - QGP interactions,
the influence of hadronization and of the final hadronic rescattering. This allowed for an understanding of this physical processes which influence the heavy quark observables.
We observed that the initial parton distribution function in large nuclei shows already to a strong suppression of low $p_T$
heavy quarks as compared to the scaled pp parton distribution function. This agrees quantitatively with calculations, which use explicitly generalized parton distribution function.
At high $p_T$, where the parton distribution functions in nucleons and heavy nuclei are similar, the interaction of the heavy quarks with the QGP medium are responsible for the low $R_{AA}$ values. The other processes are subdominant but influence quantitatively the results. The elliptic flow at
low $p_T$ is as well created by the interaction of the heavy quarks with the QGP medium but its maximum is shifted in the hadronization process. At high $p_T$ it is the path length difference which creates the finite $v_2$ values. 
Despite EPOS4HQ is based on the fundamental Gribov-Regge approach and describes all presently available heavy hadron data in heavy ion collisions it should be noted that EPOS4HQ
needs parametrization to cope with kinematic ranges, which are not accessible by more fundamental approaches. Therefore it cannot be proven whether all the subprocesses are described correctly. The agreement of the EPOS4 results with the light hadron section reduced the uncertainty considerable. An experimental study of the system size dependence would further help to reduce the uncertainties.

\vspace{1cm}
\noindent {\bf Acknowledgement}: 
We thank E. Bratkovskaya and T. Song for fruitful discussions; 
This work is supported by the European Union’s Horizon 2020 research and innovation
program under grant agreement No 824093 (STRONG-2020).

\bibliographystyle{apsrev4-1.bst}
\bibliography{Ref}
\end{document}